# Chiral topological insulators, superconductors, and other competing orders in three dimensions

Pavan Hosur, Shinsei Ryu, and Ashvin Vishwanath

*Department of Physics, University of California, Berkeley, California 94720, USA*

We discuss the proximate phases of a three-dimensional system with Dirac-like dispersion. Using the cubic lattice with plaquette $\pi$-flux as a model, we find, among other phases, a chiral topological insulator and singlet topological superconductor. While the former requires a special "chiral" symmetry, the latter is stable as long as time reversal and SU(2) spin rotation symmetry are present. These phases are characterized by stable surface Dirac fermion modes, and by an integer topological invariant in the bulk. The key features of these phases are readily understood in a two dimensional limit with an appropriate pairing of Dirac nodes between layers. This Dirac node-pairing picture is also shown to apply to $\mathbb{Z}_2$ topological insulators protected by time-reversal symmetry. The nature of pointlike topological defects in these phases is also investigated, revealing an interesting duality relation between these topological phases and the Neel phase.

     

## I. INTRODUCTION

The experimental discovery of graphene[1] has led to an explosion of interest in semi-metals with Dirac dispersion. One of the remarkable properties of a Dirac semimetal is its proximity to a variety of orders, which when established, lead to an energy gap. In the context of graphene, charge density wave[2,3] and valence bond solid (VBS) (or Kekule) order[4–6] as well as antiferromagnetism[7–9] are known to induce a gap, and lead to an insulating state. Several years back, Haldane pointed out that the integer quantum Hall state could be realized starting from the graphene semimetal, in the absence of external magnetic fields.[10] A valuable outcome of this Dirac proximity approach, was the discovery of an entirely new phase of matter, the $\mathbb{Z}_2$ quantum spin Hall insulator,[11–13] obtained in theory by perturbing the graphene Dirac dispersion. By analogy, here we study three dimensional Dirac fermions, and their proximate gapped phases, on a cubic lattice.

In three dimensions, Dirac points naturally occur in some heavy materials such as bismuth and antimony, with strong spin-orbit (SO) interactions. A three dimensional version of the quantum spin Hall state—the $\mathbb{Z}_2$ topological insulator,[14–17] can be realized by appropriately perturbing such a state, as demonstrated in Ref. 14, in a toy model on the diamond lattice. According to recent experiments, this phase is believed to be realized by several Bi-based materials including $Bi_{0.9}Sb_{0.1}$,[18] $Bi_2Se_3$,[19] and $Bi_2Te_3$.[20] Both the $\mathbb{Z}_2$ quantum spin Hall and the $\mathbb{Z}_2$ topological insulator phases require time-reversal symmetry (TRS) to be preserved. The $\mathbb{Z}_2$ index represents the fact that only an odd number of edge or surface Dirac nodes are stable in these phases.

In contrast, in this paper, we study a toy model on the cubic lattice, with $\pi$ flux through the faces, which realizes three dimensional Dirac fermions, and identify the proximate states. To begin with, we consider insulating phases of spin polarized electrons. In addition to conventional insulators, e.g., with charge or bond order, we also find an additional topological insulator phase within this model, the chiral topological insulator (cTI). This provides a concrete realization of this phase, which was recently predicted on the basis of a general topological classification of three dimensional insulators in different symmetry classes.[21,22] This phase is distinct from the spin-orbit $\mathbb{Z}_2$ topological insulators in two main respects. First, it is realized in the *absence* of time-reversal symmetry. Instead, it relies on another discrete symmetry called the chiral symmetry. Second, these insulators also host protected Dirac nodes at their surface, but *any integer* number of Dirac nodes is stable on its surface. Thus, it has a $\mathbb{Z}$ rather than $\mathbb{Z}_2$ character. In an insulator, chiral symmetry restricts us to Hamiltonians with only hopping terms between opposite sublattices. Clearly, this is not a physical symmetry, and hence, such insulators are less robust than topological insulators protected by time reversal symmetry. However, our results will be relevant if such symmetry breaking terms are weak, or in engineered band structures in lattice cold atom systems.[23] With spin, an interesting gapped state that can be reached from the Dirac limit is the singlet topological superconductor (sTS), also first discussed in Ref. 21. This state also possesses protected Dirac surface states. The stability of these states is guaranteed, as long as time reversal symmetry and SU(2) spin symmetry, both physical symmetries, are preserved.

The Dirac limit allows for an easy calculation of the charge (spin) response of the cTI (sTS), and provides an intuitive picture of these phases. For example, the cTI can be understood as arising from a quasi 2D limit of layered Dirac semimetals, with a particular pattern of node pairings, leading to a bulk gap, but protected surface states. It is hoped that this intuition will help in the search for realistic examples of these phases. Additionally, this picture helps in understanding $\mathbb{Z}_2$ topological insulators protected by TRS whose bulk Dirac nodes are at time-reversal-invariant momenta (TRIM), such as $Bi_2Se_3$, $Bi_2Te_3$, and $Sb_2Se_3$.[24]

Finally, we utilize the Dirac starting point to derive relations between different gapped phases. We show that there is a duality between Neel and VBS phases: point defects of the Neel order (hedgehogs) are found to carry quantum numbers of the VBS state and vice versa. This is done by studying the midgap states induced by these defects, and the results agree with spin model calculations[25] that are appropriate deep in the insulating limit. Thus, the Dirac approach is a convenient way to capture universal properties of the gapped phases in its vicinity. These results are also derived following a technique applied to the one and two dimensional cases,[26–29] by integrating out the Dirac fermions and deriving an effective action for a set of orders. In particular, we focus on the Berry's phase (or Wess-Zumino-Witten) term which, when present, implies nontrivial quantum interference between them. Such sets of "quantum competing" orders can be readily identified within this formalism. We show that in addition to Neel and VBS orders, interestingly, Neel order and the singlet topological superconductor also share such a relation. The consequence of such a relation in three dimensions (3D) is discussed.

The organization of this paper is as follows. We introduce the cubic lattice Dirac model and a transformation on the low energy Dirac fermions to bring it into a "normal" form in Sec. II B, from which one can easily read off the orders that lead to an energy gap. The chiral topological insulator is identified, and a microscopic model with hopping along the body diagonals of the cube, is shown to lead to this phase. An intuitive picture in terms of a quasi 2D starting point is developed in Sec. III B, which directly demonstrates the existence of surface Dirac states. Here, we also show how a $\mathbb{Z}_2$ topological insulator protected by TRS with bulk Dirac nodes at TRIM can be understood within this picture. In Sec. IV the magnetoelectric coefficient "$\theta$" of such an insulator is argued to be quantized, and is calculated to be $\theta = \pi$ for the spinless fermion model we discuss. Introducing spin, and studying gapped superconducting states, we show in Sec. VI that only a pair of singlet superconductors is allowed, which, in addition to the regular single $s$-wave paired state, includes a singlet topological superconductor, with pairing along the body diagonals. Section VII describes an attempt to move toward a more physical realization of the cTI phase, utilizing a layered honeycomb lattice structure. Finally, in Sec. VIII, we explore some topological properties of the 3D Dirac fermion system by studying its physics in the presence of point topological defects in its order parameters, and deriving the Berry's phase terms that determine quantum interference of different orders.

A word on our notation is warranted before taking the plunge into the main content. Throughout the paper, we use "$\mathbb{H}$" to denote the full Hamiltonian for a system. An ordinary "$H$" shall represent the Hamiltonian as a function of certain indices, the expression being valid over the entire energy range, and a calligraphic "$\mathcal{H}$" shall represent the Hamiltonian at low energies. For example, $\mathbb{H} = \sum_\mathbf{k} \psi_\mathbf{k}^\dagger H_\mathbf{k} \psi_\mathbf{k}$ when momentum is conserved and $H_{\mathbf{Q}+\mathbf{k}} \simeq \mathcal{H}_\mathbf{k}$ for small $\mathbf{k}$, if $H_\mathbf{Q} = 0$.

## II. CUBIC LATTICE DIRAC MODEL

Consider a 3D tight-binding model of *spinless* fermions on the cubic lattice shown in Fig. 1,

$$\mathbb{H}_0 = - \sum_{\langle R, R' \rangle} t_{RR'} c_R^\dagger c_{R'}, \qquad (1)$$

where $c_R$ is the fermion annihilation operator at site $R$ on the cubic lattice and the nearest-neighbor hoppings are chosen so that each square plaquette encloses $\pi$-flux ($\Pi_\square \frac{t_{RR'}}{|t_{RR'}|} = -1$). A particular gauge choice is shown in the figure, where the blue (bold) lines represent hopping with $-|t|$ and the others with $+|t|$. We choose an enlarged eight site unit cell, that turns out to be convenient for what follows, and label the sites $A_1, A_2, B_1, B_2$ in a layer and $A_{1'}, A_{2'}, B_{1'}, B_{2'}$ in the following layer. The $A$ and $B$ labels represent the two sublattices of the cubic lattice, and will be denoted by the eigenvalues of the $\tau_z$ operator, a Pauli matrix. Similarly, the 1,2 index will correspond to the $\nu_z$ operator, and the bilayer index, to the $\mu_z$ operator.

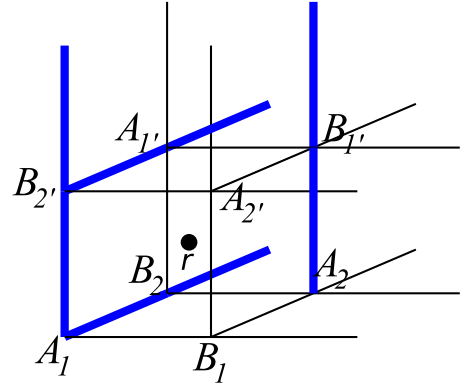

FIG. 1. (Color online) The cubic lattice with $\pi$-flux for each plaquette. Blue (bold) lines represent negative hopping integrals. **r** is at the center of cube drawn above.

With the Fourier transformation $A_{\mathbf{r}-(\mathbf{x}/2)-(\mathbf{y}/2)-(\mathbf{z}/2)} = V_A^{-1/2} \sum_\mathbf{k} e^{i\mathbf{k}\cdot[\mathbf{r}-(\mathbf{x}/2)-(\mathbf{y}/2)-(\mathbf{z}/2)]} A_\mathbf{k}$ etc., with $V_A$ being the total number of $A$ sites, **r** denoting the locations of the 8-site unit cells and **k** being in the (reduced) Brillouin zone (Bz), $\mathbf{k} \in (-\pi/2, \pi/2]^3$, the Hamiltonian is written in momentum space as

$$\mathbb{H}_0 = \sum_\mathbf{k} f_\mathbf{k}^\dagger H_\mathbf{k}^0 f_\mathbf{k}, \qquad (2)$$

where the eight component fermion operator at momentum **k** is defined by

$$f_\mathbf{k}^\dagger = (A_{1\mathbf{k}}^\dagger, A_{2\mathbf{k}}^\dagger, A_{1'\mathbf{k}}^\dagger, A_{2'\mathbf{k}}^\dagger, B_{1\mathbf{k}}^\dagger, B_{2\mathbf{k}}^\dagger, B_{1'\mathbf{k}}^\dagger, B_{2'\mathbf{k}}^\dagger) \qquad (3)$$

and

$$H_\mathbf{k}^0 = -2|t|(\cos k_x \Gamma_x + \cos k_y \Gamma_y + \cos k_z \Gamma_z), \qquad (4)$$

where

$$\Gamma_x = \tau_x, \quad \Gamma_y = \tau_y \nu_y, \quad \Gamma_z = \tau_y \mu_y \nu_x. \qquad (5)$$

Clearly, $H_\mathbf{k}^0$ anticommutes with $\tau_z$.

$$H_\mathbf{k}^0 \tau_z = -\tau_z H_\mathbf{k}^0. \qquad (6)$$

Hamiltonians for which such an anticommuting operator is present, will be called chiral Hamiltonians. A consequence is that positive and negative energy eigenvalues will come in pairs. Note that this operation amounts to changing the sign of the wave function on all $B$ sublattice sites. Applying this to the eigenstate with energy $E$, of a hopping Hamiltonian that only connects opposite sublattices, maps it to an eigenstate with energy $-E$.

$H_{\mathbf{k}}^0$ also preserves time-reversal symmetry,

$$H_{-\mathbf{k}}^{0*} = H_{\mathbf{k}}^0. \tag{7}$$

### A. Continuum limit

The energy spectrum of $H_{\mathbf{k}}^0$ is given by

$$E_{\mathbf{k}} = \pm 2|t|\sqrt{\cos^2 k_x + \cos^2 k_y + \cos^2 k_z} \tag{8}$$

and each band is fourfold degenerate for each $\mathbf{k}$. The conduction and valence bands touch at zero energy at the BZ corner,

$$\mathbf{Q} = (\pi/2, \pi/2, \pi/2). \tag{9}$$

Hence, at half-filling, it is a Dirac semimetal, with a three dimensional Dirac point. Near this point, the Hamiltonian is Dirac-like,

$$H_{\mathbf{Q}+\mathbf{k}}^0 \simeq \mathcal{H}_{\mathbf{k}} = v_F \sum_{i=x,y,z}^3 k_i \Gamma_i, \tag{10}$$

where we have introduced the Fermi velocity $v_F = 2|t|$. From now on, we will assume $v_F = 1$.

This dispersion of the Dirac semimetal can acquire a gap, leading to an insulating state, in a variety of ways. All of these require symmetry breaking of one kind or other, leading to different orders. Some obvious insulating state that can lead to such a gap (ignoring for a moment the electron spin) are:

(1) Charge density wave (CDW) order with wave-vector $(\pi, \pi, \pi)$. This will lead to a staggered potential (SP) on the two sublattices, $\Delta H_{\mathrm{CDW}} = (-1)^R \mu$. This generates a mass term for the Dirac equation, since $\Delta H \propto \tau_z$, which anticommutes with the velocity matrices $\Gamma_i$, leading to a gap.

(2) VBS order. Staggering the hopping matrix elements also opens up a gap. For example, we pick hopping along the $x$ direction and modulate their amplitude as $t_{RR+\hat{x}} = t^0 + (-1)^{R_x} \delta t$. The relevant order is called valence bond solid, since in the extreme limit where only the stronger bonds are present, the resulting insulating state may be thought of as a "molecule" composed of pairs of sites connected by these strong bonds. Similarly, one can construct VBS orders along the $y$ and $z$ directions, leading to three different mass terms. Note, these preserve the *chiral* property of the Hamiltonian, in that it still consists only of nearest-neighbor hoppings.

(3) Layered quantum Hall effect (QHE) in the $xy$, $yz$, and $zx$ layers. In three-dimensional lattices, a stacked version of the integer quantum Hall effect occurs,[30] which, however, breaks the cubic symmetry of the lattice. These orders can be realized by selecting a plane, say, the $xy$ plane, and introducing imaginary hoppings between second neighbor sites in this plane in such a way that the reflection symmetry $m_{xy}$ is broken, but $m_{yz}$ and $m_{zx}$ are unbroken. Note, time-reversal symmetry is necessarily broken here.

It is possible to systematically list all the perturbations that describe distinct insulating phases by looking at matrices that anticommute with the velocity matrices, $\Gamma_i$s. Each anticommuting term introduces a mass gap and converts the system into a true insulator, while leaving the fourfold degeneracy of each band intact. Such an insulating phase with completely degenerate conduction and valence bands is representative of a given insulating phase. We will list all such matrices in the next section, after a convenient canonical transformation that makes the counting trivial and yields a total of eight matrices (for spinless electrons). Thus, the seven orders listed above (the VBS and QHE have degeneracies of three each) do not exhaust all possible insulating states. The remaining insulator will be found to maintain the chiral condition, but will display unusual band topology. Hence, we call it the chiral topological insulator (cTI), and discuss its properties in the following section. The Dirac mass matrices corresponding to these orders and their symmetry properties are summarized in Table II in Appendix A.

### B. Transformation to normal form

It will be useful to write the kinetic part of the Dirac Hamiltonian $H_{\mathbf{k}+\mathbf{Q}}^0$ in a form where the Dirac and flavor indices are separated out. Since the Dirac matrices in three dimensions are represented by $4 \times 4$ matrices, for the eight-dimensional representation we have here this will result in two flavors of Dirac fermions. We seek a representation where the velocity matrices are independent of the flavor index. There are several distinct transformations that achieve this, and we choose to use the following unitary transformation that selects $\mu$ as the flavor index.

$$\mathcal{H}_{\mathbf{k}}^{\mathrm{Dirac}} = U H_{\mathbf{k}+\mathbf{Q}}^0 U^\dagger, \quad U = e^{i(\pi/4)\nu_x} e^{-i(\pi/4)\tau_x} e^{i(\pi/4)\mu_y \nu_y}, \tag{11}$$

hence, giving us a new set of velocity matrices $\alpha_i$,

$$\mathcal{H}_{\mathbf{k}}^{\mathrm{Dirac}} = (k_x \alpha_x + k_y \alpha_y + k_z \alpha_z), $$

$$\alpha_x = \tau_x, \quad \alpha_y = \tau_z \nu_x, \quad \alpha_z = \tau_z \nu_z. \tag{12}$$

Note, they do not involve the $\mu_a$ Pauli matrices. Now, any mass term must anticommute with these three matrices. There are two Dirac matrices that have this property, which we call $\beta_0, \beta_5$,

$$\beta_0 = \tau_y, \quad \beta_5 = \tau_z \nu_y. \tag{13}$$

Note also, that these anticommute with one another $\{\beta_0, \beta_5\} = 0$. Any Hermitian $2 \times 2$ flavor matrix multiplying one of the matrices, will also lead to a mass term. Since there are four such flavor matrices $\{\mathbb{1}_{2 \times 2}, \mu_x, \mu_y, \mu_z\}$, we have eight mass terms in all, representing eight insulators that can be accessed from the Dirac theory. The seven orders identified earlier can now be identified with their mass matrices as in Table I. In addition we identify the eighth mass term $\beta_5 \otimes \mathbb{1}_{2 \times 2}$ as that of a type of topological insulator, which satisfies the chiral condition. The chiral condition (6) is now implemented by

$$\{H_{\mathbf{k}}, \beta_0\} = 0, \tag{14}$$

and TRS for a Hamiltonian $\mathbb{H} = \Sigma_{\mathbf{k}} f_{\mathbf{k}}^\dagger H_{\mathbf{k}} f_{\mathbf{k}}$ is preserved if

TABLE I. Mass matrices ($\mathcal{M}$) in canonical representation. The $\beta$ matrices are antisymmetric and anticommute with the symmetric $\alpha$ matrices in Eq. (12) and also with each other. The "flavor" index, $\mu$, is absent from the $\alpha$-matrices.

| | $\tau_y$ | $\beta_5$ | $\tau_z \nu_y$ |
|---|---|---|---|
| Order | $\mathcal{M}$ | Order | $\mathcal{M}$ |
| CDW | $\beta_0$ | cTI | $\beta_5$ |
| QHE$_{xy}$ | $\beta_0 \mu_x$ | VBS$_z$ | $\beta_5 \mu_x$ |
| QHE$_{yz}$ | $\beta_0 \mu_y$ | VBS$_x$ | $\beta_5 \mu_y$ |
| QHE$_{zx}$ | $\beta_0 \mu_z$ | VBS$_y$ | $\beta_5 \mu_z$ |

$$\beta_0 \beta_5 \mu_y H^*_{-\mathbf{k}} \mu_y \beta_5 \beta_0 = H_{\mathbf{k}}. \quad (15)$$

While evaluating the TRS properties of the perturbations listed in Table I, note that under $\mathbf{k} \rightarrow -\mathbf{k}$, the perturbations in the second set change sign. One advantage of writing the Hamiltonian in this form is that the commutation relationships between the various mass matrices can be determined trivially. Perturbations described by commuting mass matrices are separated by a quantum critical point. For example, a quantum phase transition is required to go from the VBS phase to the cTI phase, but not to the CDW phase. The other advantage will be seen in Sec. VI.

## III. CHIRAL TOPOLOGICAL INSULATOR: PROTECTED SURFACE STATES

There is only one mass term, which preserves the chiral symmetry [Eq. (14)] and breaks TRS [Eq. (15)], and we propose that this is a 3D topological insulator discussed in Ref. 21. We call this a chiral topological insulator (cTI). On the cubic lattice, this mass term corresponds to the purely imaginary third neighbor hopping, along the body diagonal of the cube. The corresponding microscopic Hamiltonian is given by

$$H^{b.d.} = i|t''| \sum_{\mathbf{R}} \sum_{\Delta \mathbf{R}=1}^{8} c_{\mathbf{R}}^{\dagger} \Phi_{\mathbf{R}}^{site} \Phi_{\Delta \mathbf{R}}^{dir} c_{\mathbf{R}+\Delta \mathbf{R}}, \quad (16)$$

where $\mathbf{R} = (X, Y, Z)$ labels sites on a cubic lattice with a one-site unit cell and $\Delta \mathbf{R} = (\Delta X, \Delta Y, \Delta Z) = (\pm 1, \pm 1 \pm 1)$ connects the site at $\mathbf{R}$ to its eight third neighbors. $\Phi_{\mathbf{R}}^{site}$ and $\Phi_{\Delta \mathbf{R}}^{dir}$ are phase factors depending on the site $\mathbf{R}$ and the direction of $\Delta \mathbf{R}$, respectively, as

$$\Phi_{\mathbf{R}}^{site} = (-1)^{X+Z}, \quad \Phi_{\Delta \mathbf{R}}^{dir} = \Delta X \Delta Y \Delta Z. \quad (17)$$

The result is shown for a particular $\mathbf{R}$ in Fig. 2. Note that the third neighbor hopping texture that gives a cTI depends on the gauge choice. The texture shown is appropriate for our chosen gauge, in which there are negative hoppings along $Y$ and $Z$.

This renders the total Hamiltonian $\mathbb{H} = \int d^3 k f_{\mathbf{k}}^{\dagger} (H_{\mathbf{k}}^0 + H_{\mathbf{k}}^{b.d.}) f_{\mathbf{k}}$ a member of symmetry class AIII of the Altland-Zirnbauer classification.31 A class AIII topological insulator is characterized by a nonzero topological invariant $\nu$ introduced in Ref. 21, which, in this case, is given by $\nu$

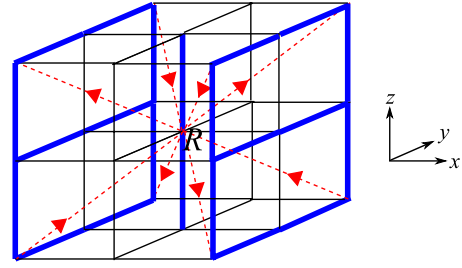

FIG. 2. (Color online) Imaginary third neighbor hopping pattern that results in a topological insulator. The arrows denote the directions in which the hopping is $+i$. The figure shows how a particular site bonds to its eight third neighbors. This pattern must be replicated around each site, after taking into account the appropriate phase factor $\Phi_{\mathbf{R}}^{site} = (-1)^{X+Z}$ for that site. In other words, all the arrows must be reversed every time the pattern is translated by a unit distance along $X$ or $Z$.

$= \pm 1$ where the sign depends on the sign of the third neighbor hopping.

### A. Protected surface states

A physical consequence of the presence of the cTI mass term with nonzero winding number $\nu = \pm 1$ is the presence of a single $(2+1)$-dimensional Dirac cone at the surface. The node is centered at zero energy. Figure 3 shows the results of a numerical calculation of the surface band structure for the (001) surface. Clearly, there is a single Dirac cone at $(\pi/2, \pi/2)$.

This is analogous to the odd number of Dirac nodes found on the surface of the spin-orbit topological insulators. However, there are several important differences. First, the surface nodes of the spin-orbit TIs are protected by TRS. In our model, however, TRS is explicitly broken, given the imaginary third neighbor hoppings and spin polarized (spinless) fermions. Instead, another discrete symmetry—the chiral symmetry—protects the node. However, a much more striking difference is the fact that the cTI can host any integer number of Dirac nodes on its surface, making it markedly different from the spin-orbit TIs which become trivial insulators if there are even numbers of Dirac nodes on their sur-faces. We will explicitly prove these features of the cTI in the ensuing sections. Also, in the cTI, the surface Dirac nodes are centered at the chemical potential, because of sublattice symmetry.

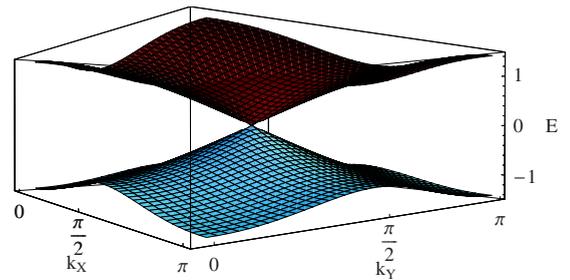

FIG. 3. (Color online) The surface spectrum in the presence of the proposed cTI term for the (001) surface for 100 bilayers in the $z$ direction. $k_x$ and $k_y$ were incremented in steps of $\pi/40$. Only the lowest conduction and highest valence band states are shown. Higher conduction and lower valence band levels that gradually merge into the bulk spectrum have not been displayed for clarity.

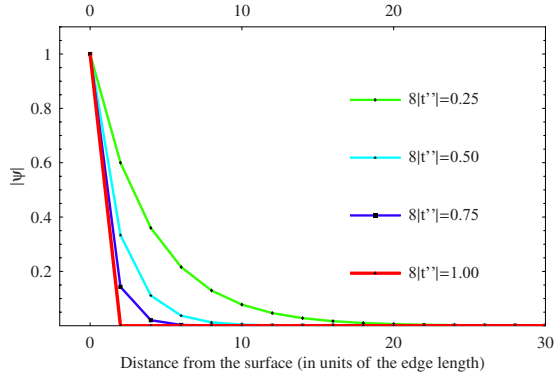

FIG. 4. (Color online) Decay of the zero-energy surface states on the $z=0$ surface into the bulk for various values of the third-neighbor hopping strength $|t''|$ (in units of the bulk Fermi velocity). The effective Dirac fermion mass is $8|t''|$. The total thickness of the lattice for this calculation is the same as that in Fig. 3, viz., 200 layers or 100 bilayers. At $8|t''|=1$, there is no penetration of the surface states into the bulk. In each case, the wave function is normalized such that it is unity on the surface.

The fact that a single Dirac cone is stable should be contrasted with the corresponding phenomenon on two-dimensional systems on a lattice, where the no-go theorem prohibits a single Dirac cone, although that requires TRS. If TRS is broken, one *can* have a single Dirac node in 2D, but it is not protected against disorder. In our model, one can never localize this surface mode[32] for arbitrary strength of disorder, as far as the disorder respects the chiral symmetry [Eq. (6)]; the dc conductivity $\sigma_{xx}$ is not affected by disorder, and is always given by the dc conductivity of a clean $(2+1)$D Dirac fermion, $\sigma_{xx}=1/\pi$ (in unit of $e^2/h$), for arbitrary strength of disorder. The (surface) density of states exhibits power-law $\rho(E)\sim|E|^{\alpha}$ with continuously varying exponent.[32]

Figure 4 shows the exponential decay of the surface states into the bulk for various values of the third-neighbor hopping strength $|t''|$. The fermion "mass" in the low-energy theory is $8|t''|$. Note, when $8|t''|=1$ the wave function is exactly localized within one unit cell of the surface. This should not come as a complete surprise. Other systems are known in condensed matter physics that have edge states bound to a single cell on the surface at a certain point in the parameter space of the coupling constants, and which decay into the bulk as we move away from this point. A famous example is the spin-1 Heisenberg chain with a biquadratic coupling, which carries a free spin-$\frac{1}{2}$ state at each end. Precisely at the "AKLT point," each of these states lives exactly on the lattice site at its end, and seeps-in along the chain as we move away from that point.[33]

### B. Physical picture

At the microscopic level, this feature of the surface spectrum can be understood by starting from a quasi-2D limit and then increasing the strength of the interaction between adjacent sheets. In the decoupled limit, each layer has a pair of Dirac nodes. In the presence of interactions, the degeneracy between the states in adjacent sheets gets destroyed. However, we will show that, on any given layer, the combination of direct and body-diagonal hopping will cause one of the nodes to interact only with the layer above and the other, only with the layer below. This way, the bulk will get gapped out but a single Dirac node will be left on the surface. For calculational convenience and in order to interpret our results most transparently, we work in a different gauge in this section.

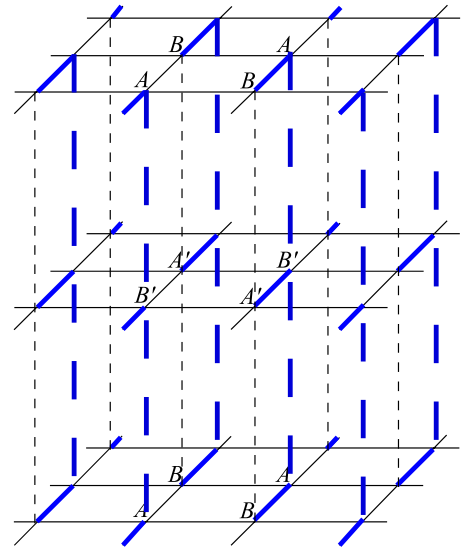

FIG. 5. (Color online) A quasi-2D approach to creating the $\pi$-flux cubic lattice. The blue (bold) lines denote negative bonds. The in-plane hoppings have strength $|t|$ and the interlayer hoppings that are shown have strength $|t_z|$. The imaginary third neighbor hoppings (strength $|t''|$, see text) have not been shown for clarity. Alternate layers are labeled by primed variables, effectively doubling the unit cell in the vertical direction.

We start by considering a system of decoupled square lattices with $\pi$-flux plaquettes (see Fig. 5). For a two-site unit cell, this has two Dirac nodes at $\mathbf{Q}_{R(L)}=[\frac{\pi}{2}(-\frac{\pi}{2}),0]$ on each layer. Then, we weakly couple these layers through regular nearest-neighbor hopping in the $z$ direction and imaginary third-neighbor interactions with textures like in Figs. 1 and 2, respectively, but modified for the current gauge choice. This is expected to mix the nodes in different layers and open a gap. Once again, for ease of calculation and interpretation of the results, we double the unit cell along $z$, i.e., imagine a stack of bilayers coupled weakly both internally and externally. Now, the Pauli matrices $\nu_{x,y,z}$ act on the space of Dirac nodes. $\tau_{x,y,z}$ act on the $A/B$ sublattice index and $\mu_{x,y,z}$ act on the bilayer index along $z$ (primed vs unprimed fields), as before.

In the basis,

$$\Psi = (A_R A_R' B_R B_R' A_L A_L' B_L B_L')^T, \quad (18)$$

where the primed and the unprimed wave functions represent different layers and $A_{1R}\equiv A_1(\mathbf{Q}_R)$ etc., the intrabilayer hopping Hamiltonian takes the form,

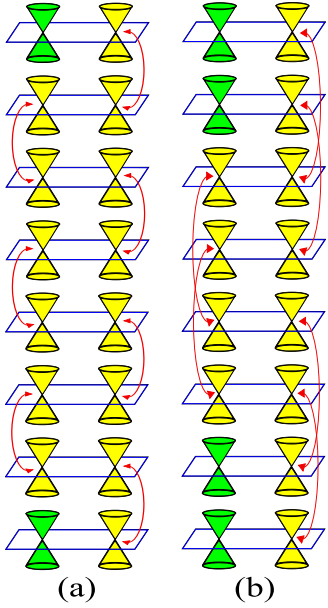

FIG. 6. (Color online) Schematic representation of the staggered interlayer mixing pattern of Dirac nodes. The degenerate Dirac points at the ends of the red arrows mix and split, opening up a gap. A surface Dirac node (colored green) is left behind in (a). A pair of surface Dirac nodes arises when we mix nodes separated by two layers as in (b).

$$\mathcal{H}_{in} = -\left(|t_z| + 4|t''|\nu_z\right)\tau_y\mu_y, \tag{19}$$

whereas hopping from $z-1$ to $z$ is described by

$$\mathcal{H}^+ = \left(|t_z| - 4|t''|\nu_z\right)i\tau_y\left(\frac{\mu_x + i\mu_y}{2}\right). \tag{20}$$

Here $|t_z|$ and $|t''|$ are the strengths of the regular and the third-neighbor hoppings, respectively.

In the limit $|t_z| = 4|t''|$, $\mathcal{H}_{in}$ and $\mathcal{H}^+$ become proportional to the orthogonal projection operators $\frac{1+\nu_z}{2}$ and $\frac{1-\nu_z}{2}$, respectively. This means that the $R$ Dirac points mix and split only *within* the bilayer, and the $L$ Dirac points, only *between* bilayers, resulting in a staggered mixing pattern as shown in Fig. 6(a). Clearly, if we terminate the lattice at a surface, a single node is left unpaired. This model is in concurrence with the numerical results presented in the previous section, because, when $|t_z|$ equals the in-plane hopping, the above limit is equivalent to the condition $8|t''| = 1$.

When $|t_z| \neq 4|t''|$, the Hamiltonian can be split into two parts: one consisting of projection operators as described above which leave the surface gapless, and the other, of the residual $t_z$, which does not open a gap anywhere (although it causes the surface states to penetrate into the bulk). Thus, the surface remains gapless for arbitrary $|t_z|$ and $|t''|$ as well. Note that since the right and the left nodes behave differently, TRS is explicitly broken as expected.

For the purists, a more rigorous calculation for the full theory shows the same.

In the basis $\Psi_{\mathbf{k}}^z = (A_{\mathbf{k}}^z A_{\mathbf{k}}^{'z} B_{\mathbf{k}}^z B_{\mathbf{k}}^{'z})^T$, the intrabilayer and interbilayer Hamiltonians take the form,

$$H_{\mathbf{k}}^{in} = -2|t|(\tau_x \cos k_x - \tau_y \sin k_y)$$
$$- \{|t_z| + 2t''[\sin(k_x + k_y) + \sin(k_x - k_y)]\}\tau_y\mu_y,$$

$$H_{\mathbf{k}}^{\pm} = \pm i\{|t_z| - 2t''[\sin(k_x + k_y) + \sin(k_x - k_y)]\}\tau_y\mu_{\pm}, \tag{21}$$

where $\mu_{\pm} = \frac{1}{2}(\mu_x \pm i\mu_y)$. By the cone-pairing picture described above, terminating the crystal from below at an unprimed layer should result in a pair of zero-energy states at the $L$ node. Indeed, solving Schrödinger's equation at $E=0$ and $\mathbf{k} = \mathbf{Q}_L$ with the boundary condition $\Psi_{\mathbf{Q}_L}^0 = 0$ gives two solutions

$$\Psi_{\mathbf{Q}_L}^z = \left(\frac{1-\alpha}{1+\alpha}\right)^{z-1}\begin{pmatrix} 1 \\ 0 \\ 0 \\ 0 \end{pmatrix},$$

$$\text{and} \quad \left(\frac{1-\alpha}{1+\alpha}\right)^{z-1}\begin{pmatrix} 0 \\ 0 \\ 1 \\ 0 \end{pmatrix}, \tag{22}$$

where $\alpha = |t_z|/|4t''|$. These are both wave functions localized at the surface. At $\mathbf{k} = \mathbf{Q}_R$, effectively $\alpha \to -\alpha$, resulting in exponentially growing solutions. When $\alpha \to 1$, $\Psi_{\mathbf{Q}_L}^z \to 0$ for $z > 1$. This is consistent with our model that when $|t_z| = 4|t''|$, there is no penetration of the surface states into the bulk.

## C. Stability of nodes and the $\mathbb{Z}$-cTI

The single Dirac node is stable against static perturbations as long as the chiral condition is preserved. This has been discussed in Ref. 21. Here, we repeat the proof in brief.

In the absence of any perturbations, the low-energy surface Hamiltonian can be written as

$$\mathcal{H}_S = -i\partial_x \eta_x - i\partial_y \eta_y = \begin{pmatrix} & -i\partial \\ -i\bar{\partial} & \end{pmatrix} \tag{23}$$

where $\eta_i$ are Pauli matrices corresponding to the two branches of the Dirac spectrum and $\partial = \partial_x - i\partial_y$, $\bar{\partial} = \partial_x + i\partial_y$. Now, since the bulk chiral operator, which is simply the CDW mass matrix, is local (i.e., purely on-site), it must have a well-defined realization on the surface too. The only matrix that anticommutes with $\mathcal{H}_S$ is $\eta_z$. Thus, this must be the chiral operator for the surface. In other words, the chiral condition on the surface is implemented by

$$\{\mathcal{H}_S, \eta_z\} = 0. \tag{24}$$

All the other perturbations which open a bulk gap are nonlocal, and thus, will couple the two Dirac cones on the two opposite surfaces of a finite slab system.

If the chiral symmetry is to be preserved, the only possible modifications to $\mathcal{H}_S$ in the presence of surface perturbations can be of the form,

$$\mathcal{H}_S \rightarrow \begin{pmatrix} & -i\,\partial + A \\ -i\bar{\partial} + A^* & \end{pmatrix}. \tag{25}$$

But this is simply a gauge transformation, and it's only effect is to shift the location of the Dirac node. Thus, the chiral symmetry protects gapless surface states.

The consequence of the chiral symmetry, though, is even more profound. Unlike the 2D and 3D time-reversal symmetric topological insulators where only an odd number of gapless surface nodes are stable leading to a $\mathbb{Z}_2$ classification, *any integer* number of surface nodes is stable on the cTI and each belongs to a distinct universality class. In the node-pairing picture, a way to generate an integer $n$ number of nodes is by assuming pairing only between nodes that are $n$ layers apart, as shown in Fig. 6(b) for $n=2$. In other words, we assume $n$ independent intercalated layers. Alternately, we can think of the $n$ layers as $n$ orbitals (or any $n$ internal degrees of freedom) on the same layer. Since the surface modes are protected by the chiral symmetry [Eq. (6)], mixing of the orbitals will not change the topological characteristics of the surface. In other words, $A$ in [Eq. (25)] becomes an $n \times n$ matrix corresponding to the orbital space and $A^* \rightarrow A^\dagger$. Diagonalizing $A$ by an appropriate similarity transformation will give $n$ copies of the single Dirac node, in general at different positions in the surface Brillouin zone. Thus, the cTI can alternately be called a "$\mathbb{Z}$-cTI" or simply, "$\mathbb{Z}$TI."

### D. Application to spin-orbit $\mathbb{Z}_2$ topological insulators

Here, we show how three-dimensional topological insulators with time reversal symmetry[14,15] can be realized using a node-pairing picture similar to the one shown above for a cTI. Since, in the above description, the left and the right Dirac nodes behave differently, TRS is explicitly broken. However, if the two nodes on the 2D sheets are both at TRIM, it is possible to realize a TRS-protected $\mathbb{Z}_2$ topological insulator (TI) through a similar mechanism. Here, we present a microscopic model for such a system.

Consider a cubic lattice with each site carrying spin-orbit coupled $s$ and $p$ orbitals with total angular momentum $\frac{1}{2}$,

$$S_+ = s\uparrow, \quad P_+ = \frac{1}{\sqrt{3}} p_0\uparrow - \sqrt{\frac{2}{3}} p_1\downarrow,$$

$$S_- = s\downarrow, \quad P_- = \frac{1}{\sqrt{3}} p_0\downarrow + \sqrt{\frac{2}{3}} p_{-1}\uparrow, \tag{26}$$

where $\uparrow(\downarrow)$ refers to up (down) spins, $s$ and $p$ are atomiclike orbitals and the subscripts on the $p$'s refer to the $z$ components of their angular momenta, i.e., $p_0 = p_z$ and $p_{\pm 1} = \frac{1}{2}(p_x \pm i p_y)$. $S_\pm (P_\pm)$ are even (odd) under inversion. Under time-reversal, they transform as

$$\mathcal{T}S_+ = S_-, \quad \mathcal{T}P_+ = P_-,$$

$$\mathcal{T}S_- = -S_+, \quad \mathcal{T}P_- = -P_+, \tag{27}$$

where $\mathcal{T}$ is the time-reversal operator.

The tight-binding Hamiltonian for this system is particularly easy to write down, if we only consider on-site and nearest-neighbor overlaps between the various orbitals. Many matrix elements vanish; for instance, the overlap integral between an $s$ and a $p_0$ orbital on the same $xy$ plane vanishes since they opposite parities under $z \rightarrow -z$. Similarly, overlap of orbitals with opposite spin vanishes. On the other hand, $s$ and $p_x$ orbitals on nearest neighbor sites in the same $xy$ plane have nonzero overlap. A similar calculation was performed in 2D in Ref. 34. The result is $\mathbb{H} = \Sigma_k \Psi_k^\dagger H_k \Psi_k$ where $\Psi^\dagger = (S_+^\dagger, S_-^\dagger, P_+^\dagger, P_-^\dagger)$ and

$$\begin{aligned}
H_k = {} & v_F (\tau_x \sigma_y \sin k_x - \tau_x \sigma_x \sin k_y + \tau_y \sin k_z) \\
& + [M + m(\cos k_x + \cos k_y + \cos k_z)]\tau_z \\
& + n(\cos k_x + \cos k_y + \cos k_z).
\end{aligned} \tag{28}$$

Here $\tau_i$ and $\sigma_i$ are Pauli matrices and $v_F$ is the Fermi velocity. $\tau_i$ act on the $S-P$ space and $\sigma_i$ act on the $\pm$ index. The first set of terms come from overlaps between orbitals of different types on neighboring sites (e.g., $S_+$ with $P_-$). These terms, being off-diagonal in the $\tau$ index are odd functions of momentum because of opposite parities of the $S$ and $P$ orbitals. The overlap between each orbital with another orbital of the same type on the same or neighboring site gives the remaining terms. Since the magnitude of the overlap integrals will in general be different for the $S$ and the $P$ orbitals, we get two kinds of terms—one, proportional to $\tau_z$ incorporates the difference in the magnitudes, and the other, proportional to identity, describes their sum. The above Hamiltonian clearly preserves TRS,

$$\sigma_y H_{-k}^* \sigma_y = H_k, \tag{29}$$

and inversion,

$$\tau_z H_{-k} \tau_z = H_k. \tag{30}$$

All the cubic symmetries are also preserved, because the basis states form representations of the full three-dimensional rotation group and the Hamiltonian preserves inversion.

The term proportional to $\tau_z$ gaps out the spectrum. The only other term that can create a gap must be proportional to $\sigma_z \tau_z$, but that breaks TRS. Therefore, the $\tau_z$ term gives a strong topological insulator (STI) for appropriate values of $m$ and $M$.

For simplicity, let us assume $n=0$. This ensures that the Fermi surface is always in the gap, unless the gap closes, in which case the Fermi surface contains the Dirac point. If $n \neq 0$, and sufficiently large, then it is possible that the system no longer remains insulating. Using the prescription outlined in Ref. 17, according to which the topological character of the band structure of an inversion symmetric system is determined by the parities of the occupied bands at the TRIM, it is straightforward to obtain the phase diagram shown in Fig. 7. In particular, in the region $\frac{1}{3} < \frac{m}{M} < 1$, a strong topological insulator is obtained. We use the notation $\nu_0; (\nu_1 \nu_2 \nu_3)$ to specify the "strong" index $\nu_0$ and the three "weak" indices $\nu_1$, $\nu_2$ and $\nu_3$ corresponding to the $x$, $y$, and $z$ directions, respectively. For an inversion symmetric system, $(-1)^{\nu_0} = \Pi_{k \in TRIM} P_k$ where $P_k$ is the product of the parity eigenval-ues of the occupied states at momentum $k$, and a STI has $\nu_0=1$. For nonzero $n$, the system is in an insulating state if $|m|,|M|>|n|$.

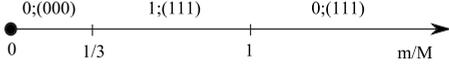

FIG. 7. Phase diagram for the 3D TRS Hamiltonian with Dirac nodes at the TRIM points as a function of the parameter $m/M$ in Eq. (28). We assume $n=0$. The numbers $\nu_0;(\nu_1\nu_2\nu_3)$ are the $\mathbb{Z}_2$ invariants that characterize the phases.[17] $\nu_0$ is the so-called "strong index" and $\nu_1$, $\nu_2$, and $\nu_3$ are the three "weak" indices corresponding to the $x$, $y$, and $z$ directions, respectively. $\nu_0=1$ for a strong topological insulator, as in the region $\frac{1}{3} < \frac{m}{M} < 1$.

To understand how the STIs are formed, it is useful to rotate the basis to $\tilde{\Psi}^\dagger = \frac{1}{\sqrt{2}}(S_+^\dagger + P_+^\dagger, S_+^\dagger + P_-^\dagger, -S_+^\dagger + P_+^\dagger, -S_-^\dagger + P_-^\dagger)$. If we turn off the interlayer coupling in the $z$-direction, we get decoupled 2D sheets with the Hamiltonian,

$$\tilde{H}_k = v_F(\tau_z\sigma_y \sin k_x - \tau_z\sigma_x \sin k_y) - [M + m(\cos k_x + \cos k_y)]\tau_x, \tag{31}$$

in the new basis. If we now turn on the interlayer coupling, the total Hamiltonian can be written as

$$\mathbb{H} = \sum_{k_x, k_y, z} (\tilde{\Psi}_z^\dagger \tilde{H}\tilde{\Psi}_z + \tilde{\Psi}_z^\dagger H^+ \tilde{\Psi}_{z-1} + \tilde{\Psi}_z^\dagger H^- \tilde{\Psi}_{z+1}), \tag{32}$$

where $H^\pm = \pm i \frac{v_F}{2} \tau_y - \frac{m}{2} \tau_x$. At the special point $m=\frac{M}{2}=v_F$, the 2D Hamiltonian is gapless at $(\pi,\pi)$ and $H^\pm=-\frac{v_F}{2}\tau_\mp$. This means that as the interlayer coupling is introduced, the Dirac points are gapped out in the following way. The first pair of time-reversal partners in each layer, $S_++P_+$ and $S_-+P_-$, mix with the second pair, $-S_++P_+$ and $-S_-+P_-$ in the layer above. Therefore, the surface of a finite insulator has a single Dirac node on the surface. At the special point chosen here, the node is exactly on the surface. If we move away from that point, the surface states start penetrating into the bulk.

In the current model, the full cubic symmetry is preserved. We may break some symmetries by, for example, changing the numerical coefficients of the $p$-orbitals in Eq. (26). Then, there will be more parameters in the model and it is possible to get several more phases with different weak indices.

## IV. CHIRAL TOPOLOGICAL INSULATOR: ELECTROMAGNETIC RESPONSE

We now discuss a field theory, which describes the electromagnetic response of a chiral topological insulator, which can be readily calculated from the Dirac limit. This can be done by considering the fermionic path integral,

$$\int \mathcal{D}[c^\dagger, c] e^{iS[c^\dagger, c, A_\mu]} =: e^{iW_{\text{eff}}[A_\mu]}, \tag{33}$$

where the fermionic action $S[c^\dagger, c, A_\mu] = \int dt L[c^\dagger, c, A_\mu]$ is given by

$$L = \sum_{R,R'} c_R^\dagger \{(i\partial_t + A_0)\delta_{RR'} - \mathcal{H}_{RR'}[A]\}c_{R'}, \tag{34}$$

and $\mathcal{H}_{RR'}[A]$ represents a (tight-binding) Hamiltonian of a chiral topological insulator, minimally coupled to the external electromagnetic U(1) gauge field $A_\mu$.[35,36] $W_{\text{eff}}[A_\mu]$ is the effective action for the electromagnetic field, which encodes all electromagnetic responses of the system. When the system of our interest is gapped in the bulk, $W_{\text{eff}}[A_\mu]$ includes the Maxwell term with modified permeability and dielectric constant. Besides the Maxwell term, $W_{\text{eff}}[A_\mu]$ can also include the so-called theta term,[37]

$$\frac{\theta e^2}{4\pi^2\hbar} \boldsymbol{E} \cdot \boldsymbol{B} = \frac{\theta e^2}{32\pi^2\hbar} \epsilon^{\mu\nu\kappa\lambda} F_{\mu\nu} F_{\kappa\lambda}, \tag{35}$$

(Here we reinstated the Planck constant and the electric charge.) The effective field theory of the same type was discussed previously for three-dimensional $\mathbb{Z}_2$ topological insulators.[35]

While in principle $\theta$ can take any value, in the presence of chiral symmetry, the theta angle is constrained to be $\theta=0$ or $\pi$. When there is chiral symmetry, under the unitary transformation (particle-hole transformation),

$$\mathcal{C}c_R\mathcal{C}^{-1} = (-1)^R c_R^\dagger, \tag{36}$$

followed by time reversal,

$$\mathcal{T}c_R(t)\mathcal{T}^{-1} = c_R(-t), \quad \mathcal{T}i\mathcal{T}^{-1} = -i, \tag{37}$$

the fermion bilinear $\int dt \sum_{R,R'} c_R^\dagger \mathcal{H}_{RR'}[A] c_{R'}$ is left unchanged while the sign of $A_0$ is flipped, i.e., $\mathcal{T}\mathcal{C}$ sends

$$\boldsymbol{E} \to -\boldsymbol{E}, \quad \boldsymbol{B} \to \boldsymbol{B}. \tag{38}$$

Thus, $\theta$ can be mapped to $-\theta$ by $\mathcal{T}\mathcal{C}$. On the other hand, under periodic boundary conditions, the theta term is invariant under $\theta \to \theta + 2\pi$. Thus, chiral symmetry is consistent with $\theta=\pi$ as well as $\theta=0$. As we will show in Appendix B, by integrating out fermions explicitly, a chiral topological insulator realizes $\theta=\pi$; $\theta=\pi$ is a hallmark of chiral topological insulators.

This is anticipated since the theta term is a surface term,

$$\epsilon^{\mu\nu\kappa\lambda} F_{\mu\nu} F_{\kappa\lambda} = 2\epsilon^{\mu\nu\kappa\lambda} \partial_\mu(A_\nu F_{\kappa\lambda}), \tag{39}$$

and upon partial integration gives rise to the $(2+1)$-dimensional Chern-Simons action at the surface of chiral topological insulator. When $\theta=\pi$, the Hall conductivity $\sigma_{xy}$ on the surface, which is the coefficient of the surface Chern-Simons action, is one half, $\sigma_{xy}=\pm 1/2$ (in unit of $e^2/h$),

$$S_{\text{CS}} = \frac{\sigma_{xy}}{4\pi} \int d^3x \epsilon^{\mu\nu\rho} A_\mu \partial_\nu A_\rho, \quad \sigma_{xy} = \pm\frac{1}{2}, \tag{40}$$

which indeed reproduces our results from microscopic calculations.

Note that the effect of the theta term shows up only when $\theta$ is nonuniform.[35,38] In particular, when we make an interface between a chiral topological insulator and a trivial insulator (or vacuum), the theta angle changes from $\theta=\pi$ (inside the topological insulator) to $\theta=0$ (outside the topological in-sulator). Then, in the presence of the chiral symmetry, there exist gapless surface states. If the chiral symmetry is broken on the surface, for example, by breaking the $A/B$ sublattice symmetry, the surface states will be gapped and lead to a quantized hall response. This also defines a smooth path for $\theta$ to evolve from 0 to $\pi$ near the surface.

## V. CHIRAL TOPOLOGICAL INSULATOR WITH SPIN

In order to explore all possible insulating and superconducting phases, it is necessary to introduce spin using the spin Pauli matrices $\sigma_i$,

$$\vec{\mathcal{M}}_{spinful} = \mathcal{M}_{spinless} \otimes \vec{\sigma}, \tag{41}$$

For example, the tensor product of the CDW mass matrix with the spin Pauli matrices gives a mass matrix describing $(\pi, \pi, \pi)$ Neel order. Similarly, we can get a spin-chiral topological insulator (s-cTI) by starting from the cTI mass ($\beta_5$ in Table I) and considering its tensor product with $\sigma_{x,y,z}$, forming a spin dependent and TRS mass term. The resulting Hamiltonian is $\mathbb{H}_{\text{s-cTI}} = \Sigma_{\mathbf{k}}(f_{\mathbf{k}\uparrow}^\dagger, f_{\mathbf{k}\downarrow}^\dagger)\sigma_y \otimes H_{\mathbf{k}}^{cti}(f_{\mathbf{k}\uparrow}, f_{\mathbf{k}\downarrow})^T$ where

$$\begin{aligned} H_{\mathbf{k}}^{cti} = 2|t''|[\sin(k_x + k_y - k_z) - \sin(k_x + k_y - k_z) \\ + \sin(k_x - k_y + k_z) - \sin(k_x - k_y - k_z)]\tau_y\mu_y\nu_z \end{aligned} \tag{42}$$

and $f_{\mathbf{k}\uparrow(\downarrow)}$ is as defined in Eq. (3) with obvious extension to include spin. According to Sec. III, there are now two surface massless Dirac fermion states, one for each spin. The gapless nature of these surface modes turns out to be stable, and the gapped Hamiltonian $H_{\text{s-cTI}}$ is indeed a topological insulator. In the terminology of Ref. 21, they can be also called class CII topological insulator. In general, when arbitrary spinorbit interactions are permitted, spin chiral topological insulators are characterized by a $\mathbb{Z}_2$ quantity rather than the integer winding number, which is the even-odd parity of the winding number $\nu$ for either one of the two spin sectors. Spin chiral topological insulators are in many ways analogous to the more familiar quantum spin Hall effect (QSHE) in two spatial dimensions, but require the chiral symmetry in addition to TRS. To have an intuitive understanding of the QSHE, one can first start from two decoupled and independent QHE states with opposite chirality for each spin and glue them together. More general QSH states can then be obtained by rotating the $S_z$ conserving QSHE by SU(2) rotation, which is quite analogous to the construction of the spin chiral topological insulators above.

## VI. SINGLET TOPOLOGICAL SUPERCONDUCTORS

The $\pi$-flux cubic lattice can also host various kinds of superconducting orders, which can be obtained by performing particle-hole transformations on the spin versions of the insulators. In order to enumerate the various classes of proximate superconductors, it is sufficient to look at the low-energy physics in the vicinity of the Dirac nodes. Assuming pairing between opposite crystal momenta, and looking for fully gapped superconductors, we may write a general low-energy Hamiltonian as

$$\begin{aligned} H(\mathbf{Q} + \mathbf{k}) &\simeq \mathcal{H}_{SC}(\mathbf{k}) \\ &= \frac{1}{2}\sum_{\mathbf{k}}(F_{\mathbf{k}}^\dagger, F_{-\mathbf{k}}) \\ &\times \begin{bmatrix} (\mathbf{k}\cdot\boldsymbol{\alpha}\otimes\mathbb{1}_{4\times 4}) & \Delta \\ \Delta^\dagger & (\mathbf{k}\cdot\boldsymbol{\alpha}\otimes\mathbb{1}_{4\times 4}) \end{bmatrix}\begin{pmatrix} F_{\mathbf{k}} \\ F_{-\mathbf{k}}^\dagger \end{pmatrix}, \end{aligned} \tag{43}$$

where $F_{\mathbf{k}}$ is a fermion operator with 16 components, including the 8 sublattice indices and spin. It is related to the microscopic fermion operators $f$ via $F_n = (U \otimes \sigma_0)_{nm}f_m$, with $U$ being some unitary matrix that brings the Dirac theory into canonical form, and $\sigma_0$ the identity matrix in the spin basis. $\Delta$ is a $16\times 16$ matrix describing the pair potential in the vicinity of the Dirac node. From fermion anticommutation, it must be antisymmetric $\Delta = -\Delta^T$. Also, in order to anticommute with the kinetic energy and open up a full gap, it must be proportional to one of the two Dirac mass terms $\beta_0, \beta_5$. Thus, in general $\Delta = \beta_{0,5}\mu_i\sigma_j, i, j = 0, 1, 2, 3$ where $\mu$ is the node index. Since in the normal form described in Sec. II B, $\beta_0$ and $\beta_5$ are antisymmetric, the product $\mu_i\sigma_j$ must be symmetric. Of the 10 symmetric possibilities for this product, 9 are also symmetric in spin ($\mu_{0,x,z}\sigma_{0,x,z}$) and thus, describe triplet superconductors, whereas the lone spin antisymmetric matrix $\mu_y\sigma_y$ describes spin singlets. Therefore, our model contains two spin singlet superconductors—$\beta_0\mu_y\sigma_y$ and $\beta_5\mu_y\sigma_y$. The former is readily shown to be onsite $s$-wave pairing. The latter however is interesting and corresponds to a singlet topological superconductor (sTS), and that is what we shall focus on now.

At the microscopic level, the singlet topological superconductor arises when pairing is added along the body diagonals of the cube, in a specified form. It can be conveniently obtained from the spin chiral topological insulator $H_{\text{s-cTI}}$, by performing a spin-dependent particle-hole transformation,

$$f_{\uparrow} \rightarrow f_{\uparrow}, \quad f_{\downarrow} \rightarrow \tau_z f_{\downarrow}^\dagger, \tag{44}$$

on the spin-cTI Hamiltonian. As a result of the hopping being imaginary, the sTS is a spin singlet,

$$i|t''|A^\dagger B + \text{H.c.} \quad \text{(spinless insulator)}$$

$$\rightarrow i|t''|(A_\uparrow^\dagger A_\downarrow^\dagger)\sigma_y\begin{pmatrix} B_\uparrow \\ B_\downarrow \end{pmatrix} + \text{H.c.} \quad \text{(spin-insulator)}, \tag{45}$$

$$\rightarrow -|t''|(A_\uparrow^\dagger B_\downarrow^\dagger - A_\downarrow^\dagger B_\uparrow^\dagger) + \text{H.c.} \quad \text{(singlet SC)}. \tag{46}$$

In momentum space, Eq. (44) corresponds to

$$f_{\mathbf{k}\uparrow} \rightarrow f_{\mathbf{k}\uparrow}, \quad f_{\mathbf{k}\downarrow} \rightarrow \tau_z f_{-\mathbf{k}\downarrow}^\dagger. \tag{47}$$

This converts $H_{\text{s-cTI}}$ into a pairing Hamiltonian,

$$\begin{aligned} \mathbb{H}_{\text{sTS}} &= \sum (f_{\mathbf{k}\uparrow}^\dagger\tau_z f_{-\mathbf{k}\downarrow})\sigma_y \otimes H_{\mathbf{k}}^{cti}\begin{pmatrix} f_{\mathbf{k}\uparrow} \\ \tau_z f_{-\mathbf{k}\downarrow}^\dagger \end{pmatrix} \\ &= -i\sum_{\mathbf{k}}f_{\mathbf{k}\uparrow}^\dagger H_{\mathbf{k}}^{cti}\tau_z f_{-\mathbf{k}\downarrow}^\dagger + \text{H.c.} \end{aligned} \tag{48}$$

which is clearly a superconducting singlet.

$\mathbb{H}_{sTS}$ can be written as $\frac{1}{2}\Sigma_{\mathbf{k}}\Psi_{\mathbf{k}}^{\dagger}H_{\mathbf{k}}^{sTS}\Psi_{\mathbf{k}}$ where $\Psi_{\mathbf{k}}^{\dagger}=(f_{\mathbf{k}\uparrow}^{\dagger},f_{\mathbf{k}\downarrow}^{\dagger},f_{-\mathbf{k}\uparrow},f_{-\mathbf{k}\downarrow})$ and $H_{\mathbf{k}}^{sTS}=i\pi_{y}\sigma_{y}H_{\mathbf{k}}^{cTi}\tau_{z}$ where $\pi_{y}$ is a Pauli matrix in the particle-hole basis. It is $SU(2)_{spin}$-symmetric,

$$\pi_{x}H_{-\mathbf{k}}^{sTS*}\pi_{x}=-H_{\mathbf{k}}^{sTS}, \qquad (49)$$

it also preserves TRS,

$$\sigma_{y}H_{-\mathbf{k}}^{sTS*}\sigma_{y}=H_{\mathbf{k}}^{sTS}. \qquad (50)$$

Thus, it belongs to class CI of BdG Hamiltonians according to the Altland-Zirnbauer classification.

Having been obtained from the cTI by particle-hole rotation allows us to conclude that there are protected 2D Dirac nodes on the surface of the topological superconductor. In this minimal case, a pair of Dirac nodes is present. The intuitive node pairing picture presented for cTIs, should also hold here, which suggests a route to a topological superconductor by stacking nodal two dimensional superconductors. Since the latter are commonly encountered, we hope this might help in the search for these exotic paired states.

However, as shown in Ref. 21, the combination of TRS and $SU(2)_{spin}$-symmetry results in a *second* chiral condition (i.e., chiral symmetry, which is not related to sublattice symmetry), which in our representation is

$$\pi_{x}\sigma_{y}H^{sTS}\sigma_{y}\pi_{x}=-H^{sTS}. \qquad (51)$$

As a consequence, the surface states of the sTS are protected by the physical symmetries of time reversal and spin rotation, and are therefore robust against the destruction of the chiral symmetry [Eq. (6)], which stabilizes the cTI.

The surface states can be detected, for example, by a tunneling experiment. In the absence of disorder, the surface density of states $\rho(E)$ is linear in energy, $\rho(E)\sim|E|$, characteristic to the 2D Dirac dispersion. On the other hand, randomness is a relevant perturbation to the surface modes, and flows to a strong coupling renormalization group fixed point. The random $SU(2)$ gauge potential, known not to be able to localize the Dirac fermions, renormalizes to an exactly solved strong coupling renormalization group (RG) fixed point at long distances. Likewise to the surface Dirac fermion mode of a chiral topological insulator (see Sec. III A), disorder is not able to localize the surface Dirac fermions in a singlet topological superconductor. However, under the influence of disorder, the (tunneling) density of states $\rho(E)$ changes from a linear dependence to $\rho(E)\sim|E|^{1/7}$.[39–41]

## VII. TOWARD PHYSICAL REALIZATION

The $\pi$-flux cubic lattice, although very convenient for computational purposes, is somewhat unnatural. Here, we discuss an alternate system which has the same salient features and hence gives rise to the same physics for the ordered state as the $\pi$-flux cubic lattice, but is somewhat more physical.

A stack of honeycomb sheets coupled in such a way that every vertical face encloses a flux of $\pi$ has three dimensional Dirac nodes in its band structure. However, unlike the cubic lattice, the $\pi$-flux can be generated very naturally through SO interactions by placing an atom with strong SO interac-tions at the center of every alternate vertical face, as shown in Fig. 8.

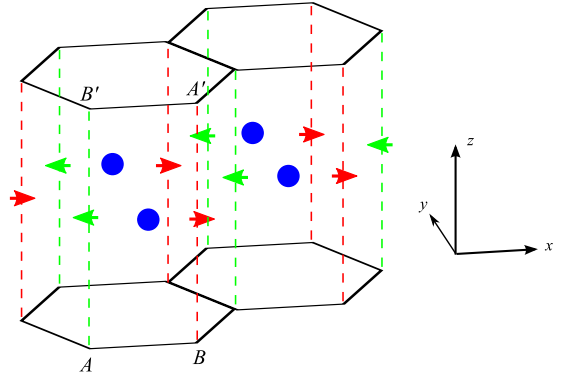

FIG. 8. (Color online) Honeycomb sheets coupled via spin-orbit interactions as a result of the electric fields (small arrows) generated by additional atoms (blue dots) at the centers of the vertical faces normal to $\hat{y}$. An electron "going up" along a green ($AB'$ bond) or red ($A'B$ bond) dotted line effectively feels a hopping of $-i\lambda_{SO}$ or $i\lambda_{SO}$, respectively. Thus, the path $AB'A'BA$ encloses a flux of $\pi$.

For a single honeycomb layer, the Hamiltonian takes the form,

$$H_{\mathbf{k}}^{honey}=-t[A_{\mathbf{k}}^{\dagger}B_{\mathbf{k}}(e^{ik_{x}}+e^{i[(-k_{x}+\sqrt{3}k_{y})/2]}+e^{i[(-k_{x}-\sqrt{3}k_{y})/2]})]+\text{H.c.}, \qquad (52)$$

$t$ being the hopping strength. For the layered system shown in Fig. 8, we must add a term,

$$\mathbb{H}_{SO}=\sum_{z}(v_{z}\hat{z}\times E\hat{x})\cdot\vec{\sigma}\equiv i|\lambda_{SO}|\sigma_{y}, \qquad (53)$$

at each $(k_{x},k_{y})$ to the sum of single layer Hamiltonians of the form Eq. (52) for each layer. Here, the velocity operator for the $z$-th layer is given by

$$v_{z}=-iA_{z}^{\dagger}(B_{z+1}'-B_{z-1}')-B_{z}^{\dagger}(A_{z+1}'-A_{z-1}')+\text{H.c.}, \qquad (54)$$

the electric field $E=-|E|$ for the $A-B'$ bonds and $+|E|$ for the $A'-B$ bonds, and $\sigma_{i}$ are Pauli matrices in the spin space. The planar momentum has been suppressed in Eq. (54) to enhance its readability. The band structure of this system has three-dimensional Dirac nodes at $\mathbf{Q}_{R(L)}=(0,\pm\frac{4\pi}{3\sqrt{3}},0)$ and the low-energy Hamiltonian now takes the form,

$$\mathcal{H}_{\mathbf{p}}=\tau_{y}\mu_{z}p_{x}+\tau_{x}\nu_{z}p_{y}-2|\lambda_{SO}|\tau_{y}\mu_{y}\sigma_{y}p_{z}, \qquad (55)$$

where $\tau$, $\nu$ and $\mu$ are Pauli matrices on the sublattice, node and layer space, respectively. This Hamiltonian preserves the chiral symmetry,

$$\tau_{z}\mathcal{H}_{\mathbf{p}}\tau_{z}=-\mathcal{H}_{\mathbf{p}}, \qquad (56)$$

and TRS,

$$\sigma_{y}\nu_{x}\mathcal{H}_{-\mathbf{p}}^{*}\nu_{x}\sigma_{y}=\mathcal{H}_{\mathbf{p}}. \qquad (57)$$

The cTI is now realized through a texture of *real* third-neighbor hoppings as shown in Fig. 9. In the low-energy theory, it corresponds to the mass term,

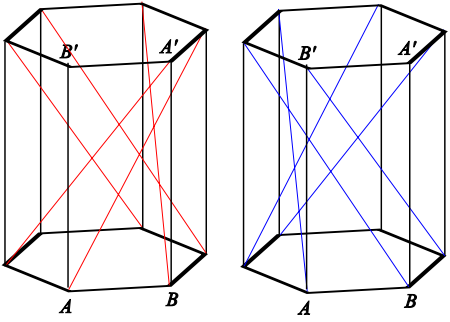

FIG. 9. (Color online) Hopping texture that results in a chiral topological insulator. The red (left figure) and blue lines (right figure) represent bonds of opposite signs when the vertical hopping is purely due to spin-orbit coupling, respectively. When ordinary real hopping in the $z$ direction is also present, the two colors can be interpreted as bonds of different strengths for some values of the parameters. Two separate figures have been drawn for clarity.

$$\mathcal{M}_{cTI} = \tau_y \nu_z \mu_x. \qquad (58)$$

This preserves TRS for spinful electrons and thus, results in the spin-cTI discussed in Sec. V. For spin-polarized electrons, the Hamiltonian for each spin individually breaks TRS, but time-reversal connects the two components of the spin. Therefore, we get two time-reversal related copies of the spinless cTI. A calculation like that in Sec. III B gives surface states that decay into the bulk as $\left(\frac{|\lambda_{SO}| - 3\sqrt{3}|t_{cTI}|}{|\lambda_{SO}| + 3\sqrt{3}|t_{cTI}|}\right)^{|z|}$, where $|t_{cTI}|$ is the strength of the third-neighbor hopping and $|z|$ is the distance from the surface ($|z|=0$ represents the surface bilayer).

This is still not very realistic, though. In a real system, we expect ordinary hopping in the $z$ direction. Also, it is more natural for a system to have alternately strong and weak third-neighbor bonds rather than bonds of opposite signs. Thus, we must include two more terms into the Hamiltonian, vertical hopping of strength $|t_z|$ and a mean third-neighbor hopping $|t_{avg}|$. $|t_{cTI}|$ now becomes the deviation from this mean. It is straightforward to check that both these terms are proportional to $\tau_x \mu_x$, and hence, destroy the pure Dirac dispersion.

The system is a topological insulator if it has zero-energy surface states while the bulk is fully gapped. The latter condition can be easily checked to be true as long as $3\sqrt{3}|t_{cTI}| > |t_z - 3t_{avg}|$, while an explicit calculation shows that the zero-energy surface states exist and decay into the bulk as $\left[\frac{|t_z - 3t_{avg}| + \tau_z(|\lambda_{SO}| - 3\sqrt{3}|t_{cTI}|)i}{|t_z - 3t_{avg}| - \tau_z(|\lambda_{SO}| + 3\sqrt{3}|t_{cTI}|)i}\right]^{|z|}$, where $\tau_z$ represents the eigenvalues of $\tau_z$ Pauli matrix, for all values of $|t_{avg}|$ and $|t_{cTI}|$. If $|t_{avg}| > |t_{cTI}|$, the bonds of opposite signs now become bonds of different strengths. Thus, even in the presence of $|t_{avg}|$ and $|t_{cTI}|$, there is a region in parameter space where the system is still a cTI.

## VIII. TOPOLOGICAL DEFECTS AND DUALITIES

Finally, we make a slight digression and discuss duality relationships between the various order parameters. Because of the plethora of phases possible for the 3D Dirac fermion system, the physics of such a system in the presence of topological defects in one or more of these phases is extremely rich. When a set of six order parameters is chosen so as to form an $O(6)$ vector in the space of mass matrices (see below for details), then the core of a point topological defect in the 3D vector field defined by three components of such a vector carries quantum numbers corresponding to the other three components. Such an intimate connection among seemingly different order parameters, which are not related by symmetry or symmetry breaking, is the heart of the non-Landau-Ginzburg transition that has been discussed in two dimensions.[28,29,42,43] In this section, we will explore the duality relationships among order parameters we have discussed so far, for the simplest physically interesting case.

### A. VBS and Neel

We start by describing the duality between the VBS and Neel order parameters.[28,29] Suppose the Hamiltonian contains VBS order,

$$\mathcal{H}(\boldsymbol{r}) = -i\partial_i \alpha_i + \beta_5 \boldsymbol{V} \cdot \boldsymbol{\mu}, \qquad (59)$$

where $\beta_5 \boldsymbol{\mu}$ are the mass terms representing three VBS orders (Table I), and $\boldsymbol{V} = (V_x, V_y, V_z)$ is the corresponding VBS order parameter and could in general, be slowly varying on the scale of the lattice spacing.

Let us study the physics of this system when $\boldsymbol{V}(\boldsymbol{r})$ contains a point topological defect at the origin. For simplicity, let us assume an isotropic "hedgehog" configuration, $\boldsymbol{V}(\boldsymbol{r}) = V(r)\hat{\boldsymbol{r}}$ where $V(0)=0$ to ensure analyticity. Then it is possible to analytically solve $\mathcal{H}(\boldsymbol{r})\psi(\boldsymbol{r})=0$ for the zero-energy modes $\psi(\boldsymbol{r})$. Note that in general these are mid gap states, but for the simple model we discuss here they appear precisely at zero energy. Since, $\boldsymbol{V}(\boldsymbol{r})$ has been assumed to be isotropic, we seek solutions that depend only on the magnitude of $\boldsymbol{r}$, i.e., $\psi(\boldsymbol{r}) \equiv \psi(r)$. Thus, we would like to solve,

$$\sin\theta\cos\phi[-i\alpha_x\psi'(r) + V(r)\beta_5\mu_x\psi(r)]$$
$$+ \sin\theta\sin\phi[-i\alpha_y\psi'(r) + V(r)\beta_5\mu_y\psi(r)]$$
$$+ \cos\theta[-i\alpha_z\psi'(r) + V(r)\beta_5\mu_z\psi(r)] = 0, \qquad (60)$$

where $\theta$ and $\phi$ are the usual spherical polar coordinates in real space.

The only angular dependence in this equation is through the trigonometric factors outside the parentheses. Therefore, the solution must satisfy,

$$-i\alpha_j\psi'(r) + V(r)\beta_5\mu_j\psi(r) = 0, \qquad (61)$$

for $j=x,y,z$. Clearly, it must be of the form,

$$\psi(r) = e^{-\int_0^r V(r)dr}\chi, \qquad (62)$$

where $\chi$ satisfies $i\alpha_x\beta_5\mu_x\chi = i\alpha_y\beta_5\mu_y\chi = i\alpha_z\beta_5\mu_z\chi = \chi$. (The eigenvectors $\chi$ of $i\alpha_j\beta_5\mu_j$ corresponding to eigenvalues $-1$ lead to unnormalizable exponentially growing solutions, $\psi(r) = e^{+\int_0^r V(r)dr}\chi$). Using the explicit forms of the matrices, it turns out that the only nonzero component of $\chi$ is the one corresponding to the $B_{1'}$ sites. Since our chosen texture for $\boldsymbol{V}(\boldsymbol{r})$ has unit topological charge, we expect, and will show more rigorously in the next subsection, that the solution just found is the only solution. Moreover, it ensures that any topologically equivalent configuration must carry a single zero mode at its core. Analogous calculations in $D=2$ were discussed in Refs. 5 and 44. We have not included the spin degree of freedom so far. For spinful electrons, there are two degenerate zero modes, and the texture carries spin $\frac{1}{2}$ if one of these levels is filled.

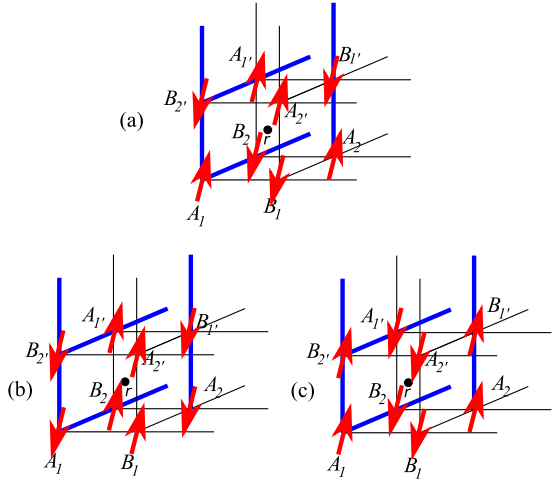

FIG. 10. (Color online) The mean field spin configuration at site $r$ (a) and the spin configurations of the two zero modes $\psi_1$ (b) and $\psi_2$ (c) at the core of a point topological defect in Neel order (See text). $\psi_1(\psi_2)$ has spins parallel (antiparallel) to the mean field spins on the $A_1 \cdot B_1 \cdot A_2 \cdot B_2$ plane and antiparallel (parallel) to them on the $A_1 B_1 A_2 B_2$ plane.

Similarly, we expect that a hedgehog in Neel order carry a "VBS-spin-$\frac{1}{2}$."[28,29] This is subtler, because the meaning of a "VBS-spin-$\frac{1}{2}$" must be defined first. This is easiest to do for a cubic lattice. Along any given cubic direction, there are clearly two degenerate ways of Kekule ordering of the bonds. These two degenerate patterns of alternating low (strong bond) and high (weak bond) energy densities are defined as the "up" and "down" components of a VBS-spin-$\frac{1}{2}$ along that direction.

We obtain the zero modes by repeating the above calculation after replacing the $\mu_i$'s above by $\sigma_i$'s and $\beta_5$ by $\beta_0$ to get the Neel order mass matrix [see Table I and Eq. (41)]. The (two) zero modes we get have the following structure: within each zero mode wave function, the real spins on neighboring sites are ferromagnetically ordered along one space direction and Neel ordered in the perpendicular plane. However, all the spins are flipped as one goes from one zero mode to the other [see Figs. 10(b) and 10(c)].

For $\vec{N} = N(r)\hat{r}$, the wave functions are

$$\psi_1(r) = e^{-\int_0^r N(r)dr} \times (0,1,0,i,i,0,1,0,i,0,-1,0,0,-1,0,i)^T,$$

$$\psi_2(r) = e^{-\int_0^r N(r)dr} \times (1,0,-i,0,0,i,0,-1,0,i,0,1,-1,0,-i,0)^T,$$

(63)

in the basis defined in Sec. II with the innermost gradation referring to spin, which is quantized along $z$. Clearly, this is related to the two VBS patterns oriented along the vertical direction of the cubic lattice. Also, under a unit translation in that direction followed by a spin flip, to maintain the mean field Neel configuration, they exchange roles as one would except from VBS orders. VBS-spins along $x$ and $y$ can be obtained by performing appropriate SU(2) rotations on the spinor $(\psi_1, \psi_2)^T$. Thus, the core of a Neel hedgehog contains states defined by VBS-spin quantum numbers.

### B. O(6) vectors of order parameters and the WZW term

We now discuss the relation between the VBS and Neel order parameters in terms of the effective field theory behind them. Subsequently, we will show that such relations exists for a much wider class of order parameters. The dual nature of the order parameters can be understood by observing that all the six matrices in the sets $\beta_5 \vec{\mu}$ (the VBS mass terms) and $\beta_0 \vec{\sigma}$ (the Neel mass terms) anticommute with each other. As a result, with the mass term $H_M = \vec{V} \cdot \beta_5 \vec{\mu} + \vec{N} \cdot \beta_5 \vec{\sigma}$ the energy eigenvalues depend only on the length of the six component vector $(\vec{V}, \vec{N})$, but not on their direction. We can separate the modulus and direction of the six components of order parameters, the latter of which forms six component vector with fixed length,

$$\hat{n} = \frac{1}{\sqrt{V^2 + N^2}}(\vec{V}, \vec{N}). \tag{64}$$

The existence of the midgap states, and the quantum numbers thereof can then be computed following the spirit of the Goldstone-Wilczek formula:[45] we integrate over gapped fermions in the presence of the slowly varying background of the $\hat{n}$ vector. The resulting effective action has a Wess-Zumino-Witten (WZW) topological term. (See Refs. 27–29 and Appendix C.) This is most conveniently written by introducing an additional fictitious coordinate $u \in [0,1]$ such that one evolves from a reference configuration at $u=0$ to the desired space-time configuration of $\hat{n}$ at $u=1$. Then,

$$S_{WZW} = i \frac{2\epsilon_{a_0 a_1 a_2 a_3 a_4 a_5}}{\pi^2} \int n^{a_0} \partial_x n^{a_1} \partial_y n^{a_2} \partial_z n^{a_3} \partial_t n^{a_4} \partial_u n^{a_5},$$

(65)

where $\epsilon$ is the antisymmetric symbol, the integral is over space-time and $u$, and a sum on the $a_i = 0 \dots 5$ is assumed.

The existence of the WZW topological term signals the fact that a solitonic configuration of the order parameter is dressed by an appropriate quantum number; if the defect is created in $V(r)[N(r)]$ it carries a quantum number related to $N(r)[V(r)]$, respectively. In turn, this is ascribed, at the microscopic level, to the existence of the midgap states in the Dirac Hamiltonian. This can be explicitly seen by assuming a static background texture for the VBS order and deriving the consequences of Eq. (65). Assuming a hedgehog defect of the VBS order, and integrating over space, one obtains an effective action for the Neel order parameter $\hat{s} = N/\|N\|$ near the core of the defect.

$$S_{WZW}^{\text{defect}} = \frac{i}{2} \int du dt \hat{s} \cdot \partial_t \hat{s} \times \partial_u \hat{s}, \quad (66)$$

which is simply the action of a spin 1/2 object. Thus a hedgehog defect of the VBS order is shown to carry spin 1/2. This also means that the hedgehog core hosts a single zero mode as claimed in the previous subsection, because then the two degenerate states of a spinor-$\frac{1}{2}$ would correspond to the zero mode being occupied or unoccupied.

Such duality relation can be found for other sets of order parameters. Among order parameters (fermion bilinears) we have discussed so far, we now look for a set of six mass matrices $M_{a=1,\ldots,6}$ which anticommutes with each other,

$$\{M_a, M_b\} = \delta_{ab}, \quad a, b = 1, \ldots, 6, \quad (67)$$

and with the Dirac kinetic term $-i\partial_i \alpha_i$, and satisfies,

$$\text{tr}[\alpha_1 \alpha_2 \alpha_3 M_1 M_2 M_3 M_4 M_5 M_6] = 16. \quad (68)$$

(See Appendix C for details.) Such order parameters are intricately correlated with each other, in the same way as the VBS and Neel order parameters are related to each other. Indeed, in Appendix C, we show for any such set of six anticommuting order parameters, the same WZW term Eq. (65) exists when the gapped fermions are integrated out.

We list below some of such 6-tuplets of order parameters for the 3D $\pi$-flux lattice model. We first note that the triplet of VBS order parameters (VBS$_{x,y,z}$) can form a pair with CDW and $s$-wave superconductivity (the real and imaginary components) (sSC). Similarly, replacing the three VBS orders with a triplet of spin chiral topological insulators (s-cTI$_{x,y,z}$) (class CII) from Sec. V also leads to a WZW term in the action,

$$\{\text{VBS}_{x,y,z}, \text{CDW}, \text{sSC}\}, \quad (69)$$

$$\{\text{s-cTI}_{x,y,z}, \text{CDW}, \text{sSC}\}. \quad (70)$$

These are generalizations of known relations in two spatial dimensions discussed in Refs. 43, 46, and 47, respectively.

Finally the sTS discussed in Sec. VI can be paired with the triplet of Neel orders, and the cTI with spin degeneracy,

$$\{\text{Neel}_{x,y,z}, \text{cTI}, \text{sTS}\}. \quad (71)$$

The last set is unique in a sense that it relates a singlet superconducting order to the triplet of Neel orders. Indeed, there is no analog of this in 2D, either on the $\pi$-flux square lattice or the honeycomb lattice, where singlet SC orders can be paired only with easy-plane SDW order (SDW$_{x,y}$), but not the full SU(2) set of Neel orders.[48] If we consider a situation where only the Neel and superconducting orders are relevant, and the sixth component above can be neglected, then one can derive a topological term for this five component set. This is the higher dimensional analog of the Haldane O(3) topological term for the spin 1/2 chain. Thus, if $\hat{n}$ is the five component unit vector comprising of three Neel and the two (real and imaginary parts) of the singlet Topological Superconductor, we have the following theta term in the action,

$$S_{\text{Top}} = i\theta Q,$$

$$Q = \frac{3}{8\pi^2} \int d^3x dt \epsilon_{a_1 a_2 a_3 a_4 a_5} n^{a_1} \partial_x n^{a_2} \partial_y n^{a_3} \partial_z n^{a_4} \partial_t n^{a_5}, \quad (72)$$

where $\theta = \pi$, and the quantity $Q$ is an integer characterizing the topology $\Pi_4[S_4]$ of smooth four dimensional space-time configurations of this five component vector. The quantum interference between Neel order and superconductivity in $D=3$ is described by an SO(5) model with a topological term. What is the physical consequence of such a term? In $D=2$, these are associated with unconventional (deconfined) quantum critical points between the pair of ordered states.[28,29,42] However, in $D=3$, a stable insulating spin liquid phase, with a spin gap, can separate the Neel and superconducting state. Such a spin liquid is expected to have electric and magnetic charges which are associated with the spin and superconducting orders. Condensing one or the other will lead to the two ordered phases. Further study of such competing orders in $D=3$ is left to future work.

## IX. CONCLUSIONS

In this paper, we have discussed *chiral* topological insulators and *singlet* topological superconductors in three spatial dimensions, proposed in Ref. 21. We constructed two concrete lattice models that realize a chiral topological insulator in symmetry class AIII. The first model is constructed by starting from the 3D $\pi$-flux cubic lattice model. The second model consists of stacked honeycomb layers with string SO interactions generating nontrivial $\pi$-flux for hopping in the direction perpendicular to the layers. While the stacked honeycomb lattice model is quasi-realistic, the 3D $\pi$-flux lattice is not particularly realistic. Nevertheless, it is a convenient canonical model to uncover interesting properties shared by general chiral topological insulators.

In many ways, a chiral topological insulator can be viewed as a close cousin of the known topological states in 3D, such as a $\mathbb{Z}_2$ topological insulator. A hallmark of both of these states is an appearance of nontrivial surface modes when topological bulk states are terminated by a boundary. However, for a chiral topological insulator, an *arbitrary* number of flavors of Dirac fermions can appear at the surface and be stable. We discussed a physically transparent picture of the chiral topological insulator, which explains the appearance of surface Dirac fermion states, and their stability in the presence of chiral symmetry. A similar picture also explains the stability of TRS $\mathbb{Z}_2$ topological insulators whose bulk Dirac nodes are centered at TRIM. It is shown that the $\theta = \pi$ axion electrodynamics can also be realized in chiral topological insulators, in addition to the known realization in $\mathbb{Z}_2$ topological insulators. We should also stress that chiral symmetry, which is realized in the models discussed here as sublattice symmetry, is likely broken in any realistic systems. Nevertheless, as far as breaking of chiral symmetry is sufficiently weak, $\theta$ is expected to be close to $\pi$, and can still have a sizable effect.

It is also worth while mentioning that chiral symmetry need not to be realized only as sublattice symmetry. Indeed, class AIII symmetry can be realized in the BdG Hamiltonians for $S_z$-conserving superconductors. Thus, chiral symmetry when realized in this way is much more robust than sublattice symmetry.

TABLE II. Mass terms and their symmetries for spinless fermions. ○ and × denote preserved and broken symmetry, respectively.

| Mass matrix ($\mathcal{M}$) | Physical interpretation | TRS (spinless) | Chiral symmetry | Inversion |
|---|---|---|---|---|
| $\tau_y \nu_z$ | VBS$_x$ | ○ | ○ | ○ |
| $-\tau_y \mu_z \nu_x$ | VBS$_y$ | ○ | ○ | ○ |
| $-\tau_y \mu_x \nu_x$ | VBS$_z$ | ○ | ○ | ○ |
| $\tau_z$ | CDW | ○ | × | × |
| $\tau_y \mu_y \nu_z$ | cTI | × | ○ | ○ |
| $\tau_z \mu_y$ | QH$_{yz}$ | × | × | × |
| $-\tau_z \mu_x \nu_y$ | QH$_{zx}$ | × | × | × |
| $-\tau_z \mu_z \nu_y$ | QH$_{xy}$ | × | × | × |

Furthermore, in the 3D $\pi$-flux lattice model with inclusion of spin degree of freedom, we found a spin-chiral topological insulator (topological insulator in class CII), and also a singlet topological superconductor (topological superconductor in class CI). The latter is stable as long as the physical symmetries of SU(2) spin rotation and time reversal are present.

Finally, utilizing the proximity to a Dirac state, we derived an interesting correlation, or "duality," between the singlet topological superconductor and Neel order. These order parameters are dual in the sense that a topological defect in either one of these phases carry complementary quantum numbers: e.g., a defect in the Neel vector ("hedgehog") can carry electric charge. We also find many such 6-tuplets of order parameters, including a six component vector consists of three Neel order and three VBS order parameters. These dualities are a natural extension of those discussed in 1D and 2D quantum spin models, the latter in the context of deconfined criticality. While in this paper we have studied the properties of these topological defects at single particle level, and hence the topological defects are static objects, we cannot resist contemplating more interesting situations where they are dynamical entities. In particular, it is interesting to ask if there is a counterpart of the non-Landau-Ginzburg transition, realized in two dimensions, that can exist, possibly in the presence of strong electron correlations in three dimensions. This is left for future study.

### ACKNOWLEDGMENTS

We thank Ying Ran and Tarun Grover for insightful discussions, and the Center for Condensed Matter Theory at University of California, Berkeley (S.R.) and NSF-DMR-0645691 (A.V.) for support.

### APPENDIX A: MASS MATRICES AND CORRESPONDING ORDERS

Table II lists the mass matrices for the various bilinears in a geometrically friendly representation, i.e., one in which the eight components of the spinor are simply the eight sites of the unit cell. The kinetic matrices are $\Gamma_x = \tau_x, \Gamma_y = \tau_y \nu_y, \Gamma_z = \tau_y \mu_y \nu_x$ (see Sec. II).

### APPENDIX B: DERIVATION OF THE THETA TERM

#### 1. Dirac model

In this appendix, we demonstrate $\theta = \pi$ in the cTI we introduced by carrying out the fermionic path integral. Below, we will use the Euclidean formalism (imaginary-time path integral), in which case the theta term appears as the imaginary part of the effective action while the Maxwell term appears as the real part. To compute the theta angle of the effective action within the continuum Dirac Hamiltonian, it will prove useful to compare the effective action for two different states, VBS and cTI,

$$W_{\text{eff,VBS}}, \quad W_{\text{eff,cTI}}. \tag{B1}$$

The two states can be connected to each other by continuous (adiabatic) deformation of the Hamiltonian, if we break the chiral symmetry during the deformation. Namely, we can interpolate, by one parameter, say, $\alpha \in [0, \pi]$, the two mass terms representing the cTI ($M_{\text{cTI}}$), and VBS ($M_{\text{VBS}}$), respectively, without closing the bulk band gap,

$$M(\alpha = 0) = M_{\text{cTI}}, \quad M(\alpha = \pi) = M_{\text{VBS}}. \tag{B2}$$

The variation of the theta angle with respect to $\alpha$, $\delta\theta/\delta\alpha$, can then be computed along the deformation path. We know that the imaginary part of $W_{\text{eff}}$ should be zero for VBS, Im $W_{\text{eff,VBS}} = 0$, since the VBS state can continuously be connected, without breaking chiral symmetry, to the trivial insulator, $\theta(\alpha) = 0$. The theta angle for cTI can then be obtained by integrating the variation $\delta\theta/\delta\alpha$ with the boundary condition $\theta(\alpha = \pi) = 0$.

We now compute the variation $\delta\theta/\delta\alpha$. To connect $M_{\text{cTI}}$ and $M_{\text{VBS}}$, we can flip the sign of the mass term for a 4 ×4 subsector of the 8×8 Dirac Hamiltonian, while keeping the other half intact: when we smoothly connect the Hamiltonians with the masses $M_{\text{cTI}}$ and $M_{\text{VBS}}$, $\mathcal{H}_{\mathbf{k}}^0 + (1-t)M_{\text{cTI}} + t M_{\text{VBS}}$ ($0 \le t \le 1$), keeping chiral symmetry during the interpolation, four out of eight eigenvalues (for each $\mathbf{k}$) cross while the remaining four are not affected by the interpolation. We thus consider the following single 4×4 continuum Dirac model,

$$\mathcal{H} = \boldsymbol{k} \cdot \boldsymbol{\alpha} + m\beta. \tag{B3}$$

The corresponding partition function and the imaginary-time action are,

$$Z = \int \mathcal{D}[\psi^\dagger, \psi] e^{-S},$$

$$S = \int d\tau d^3 r \psi^\dagger (\partial_\tau - i\,\partial \cdot \boldsymbol{\alpha} + m\beta)\psi. \tag{B4}$$

With $\bar{\psi} = \psi^\dagger \beta$, and with the inclusion of the background gauge field, the action can be written as

$$S = \int d^4 x \, \bar{\psi}[(\partial_\mu + iA_\mu)\gamma_\mu + m]\psi, \tag{B5}$$

where $\mu = 1, \ldots, 4$, and we have introduced Euclidean gamma matrices, $\gamma_{i=1,2,3} = -i\beta\alpha_i$, $\gamma_4 = \beta$, which satisfy

$$\{\gamma_\mu, \gamma_\nu\} = 2\delta_{\mu\nu}, \quad \gamma_\mu^\dagger = \gamma_\mu, \quad \mu = 1,2,3,4. \tag{B6}$$

We also introduce

$$\gamma_5 = -\gamma_1 \gamma_2 \gamma_3 \gamma_4. \tag{B7}$$

As advertised, we can flip the sign of mass, in a continuous fashion, by the following chiral rotation

$$\psi \to \psi = e^{i\alpha\gamma_5/2}\psi', \quad \bar{\psi} \to \bar{\psi} = \bar{\psi}' e^{i\alpha\gamma_5/2}, \tag{B8}$$

under which,

$$\bar{\psi}(\nabla_\mu \gamma_\mu + m)\psi = \bar{\psi}'[\nabla_\mu \gamma_\mu + m'(\alpha)]\psi',$$

$$m'(\alpha) = me^{i\alpha\gamma_5} = m[\cos\alpha + i\gamma_5 \sin\alpha], \tag{B9}$$

so that $m'(\alpha = 0) = m$ and $m'(\alpha = \pi) = -m$.

### 2. Calculation of the effective action by gradient expansion

The fermionic path integral defined by Eqs. (B4) and (B5) can be evaluated for slowly varying external gauge field $A_\mu$ by derivative expansion. Given the fact that the space-time variation of the theta angle couples to the electromagnetic field, it is convenient to consider the case where the mass term also changes slowly in space time according to,

$$m \to m\{\cos[\alpha(\tau, x)] + i\gamma_5 \sin[\alpha(\tau, x)]\}$$

$$\simeq m + i\gamma_5 \delta m(\tau, x),$$

$$\delta m(\tau, x) = m\,\delta\alpha(\tau, x). \tag{B10}$$

Below, we compute the derivation of the theta angle with respect to small change in the mass term, $\delta\theta/\delta\alpha$, by gradient expansion.

We integrate out the fermions and derive the effective action for the gauge fields $A_\mu$ and $\delta m$,

$$\int \mathcal{D}[\bar{\psi}, \psi] e^{-S} = e^{-W_{\text{eff}}[A_\mu, \delta m]}, \tag{B11}$$

by a derivative expansion,

$$W_{\text{eff}} = -\text{Tr}\ln(G_0^{-1} - V) = -\text{Tr}\ln G_0^{-1} + \sum_{n=1}^{\infty} \frac{1}{n}\text{Tr}(G_0 V)^n, \tag{B12}$$

where $G_0$ denotes the propagator of free $(3+1)D$ massive Dirac fermions, which is given in momentum space by

$$G_0(k) = -\frac{i\slashed{k} + m}{k^2 + m^2}, \tag{B13}$$

while

$$V(q) = +i\slashed{A}(q) + i\gamma_5 \delta m(q). \tag{B14}$$

The resultant effective action, to leading order in the derivative expansion, takes the following form

$$i\delta\,\text{Im}\,W_{\text{eff}} = \frac{i}{8\pi}\int d^4 x \frac{\delta m}{m}\epsilon^{\mu\nu\rho\sigma}\partial_\rho A_\mu \partial_\sigma A_\nu. \tag{B15}$$

Integrating the variation,

$$i\,\text{Im}\,W_{\text{eff}} = \frac{i}{8\pi}\int d^4 x \epsilon^{\mu\nu\rho\sigma}\partial_\rho A_\mu \partial_\sigma A_\nu. \tag{B16}$$

### 3. Calculation of the effective action by Fujikawa method

We now give an alternative derivation based on the Fujikawa method.[49] Since $M_{\text{cTI}}$ can continuously be rotated into $M_{\text{VBS}}$, one would think, naively, $W_{\text{eff,VBS}} = W_{\text{eff,cTI}}$. This is not true, however, as we have demonstrated: $W_{\text{eff,VBS}}$ and $W_{\text{eff,cTI}}$ should differ by the theta term. The reason why this naive expectation breaks down is the chiral anomaly. The chiral transformation which rotates $M_{\text{cTI}}$ continuously into $M_{\text{VBS}}$ costs the Jacobian $\mathcal{J}$ of the path integral measure,

$$\int \mathcal{D}[\bar{\psi}, \psi] e^{-S[m]} = \int \mathcal{D}[\bar{\psi}', \psi'] \mathcal{J} e^{-S[m']}. \tag{B17}$$

The chiral anomaly (the chiral Jacobian $\mathcal{J}$) is responsible for the theta term.

We now compute the Jacobian $\mathcal{J}$ explicitly, by breaking up the chiral transformation into an infinitesimal chiral rotation,

$$\psi \to \psi = U(t)\psi_t \quad \bar{\psi} \to \bar{\psi} = \bar{\psi}_t U(t),$$

$$U(t) = e^{t\Theta}, \quad t \in [0,1], \quad \Theta = i\gamma_5 \frac{\pi}{2}. \tag{B18}$$

For each step $t$

$$\bar{\psi}(\nabla_\mu \gamma_\mu + m)\psi = \bar{\psi}_t(\nabla_\mu \gamma_\mu + me^{2i(\pi/2)t\gamma_5})\psi_t \equiv \bar{\psi}_t D_t \psi_t. \tag{B19}$$

Observe that when $t$ becomes unity, we completely flip the sign of the mass term. The Jacobian is given by

$$\mathcal{J}(t) = \exp\left[-\int_0^t du\,\mathcal{V}\mathcal{V}(u)\right],$$

$$\mathcal{W}(u) = \frac{d}{du}(\ln \text{Det } D_u). \tag{B20}$$

This can be computed as

$$\mathcal{W}(u) = \lim_{\Delta u \to 0} \frac{1}{\Delta u} \text{Tr}[D_u^{-1} \Delta u (\Theta D_u + D_u \Theta)] = 2 \text{ Tr}[\Theta]. \tag{B21}$$

This expression, if naively interpreted, is divergent and should be regularized by the heat-kernel method,

$$\mathcal{W}(u) = \lim_{M^2 \to \infty} 2 \text{ Tr}[\Theta e^{-D_u^2/M^2}], \tag{B22}$$

where $M^2$ is the regulator mass, which is sent to be infinity at the end of calculations. The trace can explicitly evaluated by inserting the set of eigenstates of $D_u$ and then using the momentum basis as

$$\mathcal{W}(u) = 2 \lim_{M^2 \to \infty} \int d^4x \int \frac{d^4k}{(2\pi)^4} \text{tr}[\Theta e^{-\langle k|D_u^2|k\rangle/M^2}],$$

where $\langle k|D_u^2|k\rangle$ is the matrix elements of $D_u^2$ in the momentum basis. Noting that,

$$(\overline{\nabla})^2 = \nabla_\mu \nabla_\mu + \frac{i}{4}[\gamma_\mu, \gamma_\nu]F_{\mu\nu}, \tag{B23}$$

and keeping pieces which survive the limit $M^2 \to \infty$,

$$\mathcal{W}(u) = 2 \lim_{M^2 \to \infty} \int d^4x \int \frac{d^4k}{(2\pi)^4}$$
$$\times \text{tr}[\Theta e^{-ikx} e^{-\left(\nabla_\mu \nabla_\mu + \frac{i}{4}[\gamma_\mu, \gamma_\nu]F_{\mu\nu}\right)/M^2} e^{+ikx}]$$
$$= 2 \int d^4x \text{ tr}\left[\Theta \frac{1}{2!}\left(-\frac{i}{4}[\gamma_\mu, \gamma_\nu]F_{\mu\nu}\right)^2\right]$$
$$\times \int \frac{d^4k}{(2\pi)^4} e^{-k_\mu k_\mu}. \tag{B24}$$

Finally, noting that $\text{tr}[\gamma_5[\gamma_\mu, \gamma_\nu][\gamma_\kappa, \gamma_\lambda]] = -16\epsilon_{\mu\nu\kappa\lambda}$ and $\int d^4k/(2\pi)^4 e^{-k_\mu k_\mu} = 1/(16\pi^2)$,

$$\mathcal{W}(u) = 2i\frac{\pi}{2}\frac{-1}{32\pi^2}\epsilon_{\mu\nu\kappa\lambda}F_{\mu\nu}F_{\kappa\lambda}. \tag{B25}$$

By integrating over $u \in [0,1]$, we reproduce the previous result.

## APPENDIX C: NONLINEAR SIGMA MODEL WITH WZW TERM

The purpose of this appendix is to show the duality relation for any six-tuplets of order parameters satisfying the anticommutation relation Eq. (67), in particular for the six-tuplets discussed in Sec. VIII. We do so by showing that when gapped fermions are integrated out in the presence of the slowly varying background of the $O(6)$ vector, there is the WZW term.

### 1. Nonunitary transformation

Let us start from the Hamiltonian contains both VBS and Neel order parameters [see Eq. (59))],

$$\mathcal{H}(\boldsymbol{r}) = -i\partial_i \alpha_i + \beta_5 \boldsymbol{V} \cdot \boldsymbol{\mu} + \beta_0 \boldsymbol{N} \cdot \boldsymbol{\sigma} \tag{C1}$$

where $\boldsymbol{V}$ represents the VBS order parameter, and $\boldsymbol{N}$ the Neel vector. The imaginary-time path integral corresponding to this Hamiltonian is given by the partition function $Z = \int \mathcal{D}[\chi^\dagger, \chi] \exp(-\int d\tau d^3x \mathcal{L})$ with the Lagrangian $\mathcal{L} = \chi^\dagger(\partial_\tau + \mathcal{H}(\boldsymbol{r}))\chi$ where $\chi^\dagger$ and $\chi$ are a fermionic path integral variable. We will integrate fermions out to derive the effective action for the $O(6)$ vector $(\boldsymbol{N}, \boldsymbol{V})$, from which we will try to read off the duality relation of the order parameters. The same procedure can be repeated for any other six-tuplets of order parameters discussed in Sec. VIII. But before doing this, we will make a change of variables to transform the action into a canonical form.

To discuss all duality relations discussed in Sec. VIII in a unified fashion, let us start from a set of nine $2^4 \times 2^4$ anti-commuting Hermitian matrices,

$$\xi_i \xi_j + \xi_j \xi_i = 2\delta_{ij}, \quad i,j = 1, \ldots, 9, \tag{C2}$$

with

$$\xi_9 = -\xi_1 \xi_2 \cdots \xi_8. \tag{C3}$$

These matrices form a spinor representation of SO(9). Three out of these matrices $\xi_{1,2,3}$ can be used to form a Dirac kinetic energy, whereas the remaining six matrices can be used as a mass matrix representing an order parameter,

$$\mathcal{H}_{\mathbf{k}} = \sum_{i=1}^{3} k_i \xi_i + \sum_{a=4}^{9} m_a \xi_a, \tag{C4}$$

where $m_{a=4,\ldots,9} \in \mathbb{R}$ represents a six-component order parameter. For example, for Eq. (C1),

$$\xi_{1,2,3} = \alpha_{1,2,3}, \quad \xi_{3,4,5} = \beta_5 \mu_{1,2,3}, \quad \xi_{5,6,7} = \beta_0 \sigma_{1,2,3}, \tag{C5}$$

and

$$m_{4,5,6} = (V_x, V_y, V_z), \quad m_{7,8,9} = (N_x, N_y, N_z). \tag{C6}$$

The imaginary-time path Lagrangian is given by

$$\mathcal{L} = \chi^\dagger \left(\partial_\tau + \sum_{i=1}^{3} k_i \xi_i + \sum_{a=4}^{9} m_a \xi_a\right)\chi. \tag{C7}$$

For other 6-tuplets discussed in Sec. VIII, we can choose similarly an appropriate set of 9 matrices $\xi_i$, where the three matrices are for the Dirac kinetic term, and the remaining three for the any $O(6)$ order parameters in Sec. VIII. To this end, observe that once we consider superconducting orders we are lead to consider $2^5 \times 2^5$ mass matrices acting on sublattice indices ($\tau$), the 1 and 2 indices ($\nu$) introduced in Fig. 1, the bilayer indices ($\mu$), spin indices ($\sigma$), and particle-hole spaces ($\pi$). However, when we limit ourselves to singlet superconductivity, by making use of spin rotation symmetry, we can always reduce the dimensionality of mass matrices down to $2^4 \times 2^4$.

We now make a change of the fermionic path integral variables to transform the Lagrangian into the canonical form [see Eq. (C9) below]. To this end, we introduce,

$$\bar{\psi} := \chi^{\dagger}\xi_9, \quad \psi := \chi,$$

$$\gamma_0 := \xi_9, \quad \gamma_i := -i\xi_9\xi_i \quad (i=1,\dots,3),$$

$$\gamma_5 = -\gamma_0\gamma_1\gamma_2\gamma_3,$$

$$\Sigma_a = (\xi_5\xi_5\xi_6\xi_7\xi_8)\xi_a \quad (a=4,\dots,8),$$ (C8)

wherein the Lagrangian in terms of the new variables is given by

$$\mathcal{L} = \bar{\psi}\left(\partial_\mu\gamma_\mu + m_9 + \sum_{a=4}^{8} m_a i\gamma_5\Sigma_a\right)\psi.$$ (C9)

The merit of this change of variables is that it untangles rotations in the order parameter space and in the real space: the mass matrices ($\Sigma_{a=4,\dots,8}$) and the matrices entering in the Dirac kinetic term ($\gamma_{\mu=0,\dots,3}$) are made mutually commuting,

$$[\gamma_\mu, \Sigma_a] = 0, \quad \forall \mu, a,$$ (C10)

where $\gamma$s and $\Sigma$s form SO(4) and SO(5), respectively,

$$\gamma_\mu\gamma_\nu + \gamma_\nu\gamma_\mu = 2\delta_{\mu\nu}, \quad \mu,\nu=0,1,2,3,$$

$$\Sigma_a\Sigma_b + \Sigma_b\Sigma_a = 2\delta_{ab}, \quad a,b=4,\dots,8.$$ (C11)

Below, we will use the following notation for the order parameters and mass matrices,

$$M := m_9 + m_a i\gamma_5\Sigma^a = |M|\sum_{l=1}^{6} n_l\Upsilon^l,$$ (C12)

where the set of matrices $\Upsilon^l$, the modulus $|M|$, and the six-component unit vector $n_l$ are introduced by

$$\Upsilon^l := \{I, i\gamma_5\Sigma_4, \cdots, i\gamma_5\Sigma_8\},$$

$$|M|^2 := m_9^2 + \sum_{a=4}^{8} m_a^2,$$

$$n_l := |M|^{-1}(m_9, m_4, \cdots, m_8).$$ (C13)

For later use, we also introduce

$$\tilde{M} := m_9 - m_a i\gamma_5\Sigma_a,$$ (C14)

which satisfies $M\tilde{M} = |M|^2 I$.

## 2. Gradient expansion

So far the order parameter $m_{4,\dots,9}$ has been assumed to be static. We now consider a situation where $m_{4,\dots,9}$ changes slowly (smoothly) in space-time, $m_{4,\dots,9} \to m_{4,\dots,9}(\tau,x)$ [$M \to M(\tau,x)$]. We consider the case where length of the vector is constant, $\sum_{a=4}^{9}|m_a(\tau,x)|^2 = $ const., whereas its direction varies. I.e., the modulus $|M|$ is constant whereas the six-component unit vector $n_l$ in Eq. (C13) changes in space-time.

We now proceed to derive the effective action for the bosonic field [the set of order parameters $m_{4,\dots,9}(\tau,x)$], the O(6) nonlinear sigma model with the WZW term, following Ref. 27. The effective action $S_{\text{eff}}$ is derived by integrating over fermions,

$$e^{-S_{\text{eff}}} := \int \mathcal{D}[\bar{\psi},\psi]e^{-S} = e^{\ln\text{Det}(\gamma_\mu\partial_\mu + M)},$$

$$S_{\text{eff}} = -\text{Tr}\ln(\gamma_\mu\partial_\mu + M) =: -\text{Tr}\ln\mathcal{D}.$$ (C15)

We compute the effective action by first computing the variation $\delta S_{\text{eff}}[M]$ under a small change in the bosonic field $M(\tau,x)$, and then by recovering the full functional $S_{\text{eff}}[M]$. Taking a small variation in $M(\tau,x)$,

$$\mathcal{D} \to \mathcal{D} + \delta\mathcal{D}, \quad \delta\mathcal{D} = \delta M,$$ (C16)

the change in the effective action to the leading order in $\delta M$ is given by

$$\delta S_{\text{eff}} = -\text{Tr}[(p^2 + |M|^2 + \gamma_\mu\partial_\mu M)^{-1} \times (-\gamma_\mu\partial_\mu + \tilde{M})\delta M].$$ (C17)

Assuming the order parameter field $M$ changes smoothly in space time, we expand

$$(p^2 + |M|^2 + \gamma_\mu\partial_\mu M)^{-1}$$
$$= [1 + (p^2 + |M|^2)^{-1}\gamma_\mu\partial_\mu M]^{-1}(p^2 + |M|^2)^{-1},$$ (C18)

in terms of the derivative $\partial_\mu M$, which leads to

$$\delta S_{\text{eff}} = -\sum_n (-1)^n\text{Tr}\{[(p^2 + |M|^2)^{-1}\gamma_\mu\partial_\mu M]^n$$
$$\times (p^2 + |M|^2)^{-1}(-\gamma_\mu\partial_\mu + \tilde{M})\delta M\} = \sum_{n=0} \delta S_{\text{eff}}^{(n)}.$$ (C19)

The term, which can potentially give rise to the WZW term, is the following piece in $\delta S_{\text{eff}}^{(4)}$:

$$\delta\Gamma := -\text{Tr}\{[(p^2 + |M|^2)^{-1}\gamma_\mu\partial_\mu M]^4 \times (p^2 + |M|^2)^{-1}\tilde{M}\delta M\}.$$ (C20)

This can be written as

$$\delta\Gamma = -|M|^6\text{tr}_{16}[\gamma_{\mu_1}\Upsilon^{a_1}\gamma_{\mu_2}\Upsilon^{a_2}\gamma_{\mu_3}\Upsilon^{a_3}\gamma_{\mu_4}\Upsilon^{a_4}\tilde{\Upsilon}^b\Upsilon^c]$$
$$\times \text{tr}_k\left\{\left[\prod_{i=1}^{4}(p^2 + |M|^2)^{-1}(\partial_{\mu_i}n_{a_i})\right]n_b(\delta n_c)\right\},$$

where $\text{tr}_{16}$ represents the 16-dimensional trace, whereas $\text{tr}_k$ represents the trace over the momenta/spatial coordinates. By noting

$$\text{tr}_{16}[\gamma_{\mu_1}\Upsilon^{a_1}\gamma_{\mu_2}\Upsilon^{a_2}\gamma_{\mu_3}\Upsilon^{a_3}\gamma_{\mu_4}\Upsilon^{a_4}\tilde{\Upsilon}^b\Upsilon^c]$$
$$= 16i\epsilon_{\mu_1\mu_2\mu_3\mu_4}\epsilon_{a_1a_2a_3a_4bc},$$ (C21)

$\delta\Gamma$ is computed as

$$\delta\Gamma = -16i\epsilon_{\mu_1\mu_2\mu_3\mu_4}\epsilon_{a_1a_2a_3a_4bc}J\int d^4x\times(\partial_{\mu_1}n_{a_1})(\partial_{\mu_2}n_{a_2})$$
$$\times(\partial_{\mu_3}n_{a_3})(\partial_{\mu_4}n_{a_4})n_b(\delta n_c),\tag{C22}$$

where the integral $J$ is given by

$$J=\int\frac{d^4p}{(2\pi)^4}\frac{|M|^6}{(p^2+|M|^2)^5}=\frac{\pi}{2^3}\frac{1}{\text{Area}(S^5)}\frac{1}{4!},\tag{C23}$$

and $\text{Area}(S^d):=2\pi^{(d+1)/2}/\Gamma[(d+1)/2]$ represents the area of the $d$-dimensional unit supersphere.

Equation (C22) can be rewritten, by introducing an artificial coordinate $u\in[0,1]$, as a surface integral,

$$\delta\Gamma=\frac{2\pi i\epsilon_{\mu_1\mu_2\mu_3\mu_4}\epsilon_{a_1a_2a_3a_4bc}}{\text{Area}(S^5)4!}\int d^4x\int_0^1du$$
$$\times\partial_u[(\partial_{\mu_1}n_{a_1})(\partial_{\mu_2}n_{a_2})(\partial_{\mu_3}n_{a_3})(\partial_{\mu_4}n_{a_4})n_b(\delta n_c)],\tag{C24}$$

where we extend the integrand properly in such a way that (integrand)=0 at $u=1$, whereas at $u=0$ the integrand gives the original expression (C22). Equation (C24) is nothing but the functional derivative of the WZW functional,

$$\Gamma=\frac{2\pi i}{\text{Area}(S^5)5!}\int_{D^5}d^5x\epsilon_{\mu_1\cdots\mu_5}\epsilon_{a_1\cdots a_6}\times(\partial_{\mu_1}n_{a_1})(\partial_{\mu_2}n_{a_2})$$
$$\times(\partial_{\mu_3}n_{a_3})(\partial_{\mu_4}n_{a_4})(\partial_{\mu_5}n_{a_5})n_{a_6},\tag{C25}$$

where $dx^5=dud^4x=dud\tau d^3x$, and the integration domain $B$ is topologically equivalent to a five-dimensional disk (i.e., $\partial D^5=S^4$) with a boundary at $u=0$. We thus conclude the effective action $S_{\text{eff}}$ includes the WZW term $\Gamma$, together with the kinetic term of the SO(6) nonlinear sigma model.

We now discuss why this implies the presence of midgap modes when a defect is created in the ordered state. For concreteness, consider the six components of $\hat{n}$ above as being composed of VBS and Neel order parameters. Introduce a static hedgehog defect in the VBS order, and derive the effective action for the remaining Neel components. If we use the ansatz $\hat{n}=[\rho(r)\hat{v},\sqrt{1-\rho^2}\hat{s}(t,u)]$, where $\rho(r=0)=0$ and $\rho(r\to\infty)=1$, and $\hat{v}$ encodes the hedgehog defect, one obtains after integration:

$$S_{WZW}^{\text{defect}}=\frac{i}{2}\int dudt\hat{s}\cdot\partial_t\hat{s}\times\partial_u\hat{s}\tag{C26}$$

which, for the Neel variables, is the action of a spin 1/2 object.